\documentclass[fleqn,usenatbib]{mnras}

\usepackage{newtxtext,newtxmath}
\usepackage[T1]{fontenc}

\DeclareRobustCommand{\VAN}[3]{#2}
\let\VANthebibliography\thebibliography
\def\thebibliography{\DeclareRobustCommand{\VAN}[3]{##3}\VANthebibliography}

\usepackage{graphicx}	% Including figure files
\graphicspath{{figures/}}
\usepackage{amsmath}	% Advanced maths commands
\usepackage{comment}
\usepackage{booktabs}
\usepackage{multirow}
\usepackage{rotating}
\usepackage{float}
\usepackage{siunitx}
\usepackage{subfigure}
\usepackage{makecell} % for more vertical space in cells
\setcellgapes{5pt}
\usepackage[ruled,vlined]{algorithm2e}
\usepackage{setspace}

\title[Pulsar Candidate Classification using SGAN]{Pulsar Candidate Identification Using Semi-Supervised Generative Adversarial Networks.}

\author[V.Balakrishnan et al.]{
Vishnu Balakrishnan,$^{1}$\thanks{E-mail: vishnu@mpifr-bonn.mpg.de}
David Champion,$^{1}$
Ewan Barr,$^{1}$ Michael Kramer,$^{1}$ Rahul Sengar,$^{2}$ 
\newauthor Matthew Bailes $^{2} $
\\
$^{1}$Max-Planck-Institut fur Radioastronomie, Auf dem Hügel 69, D-53121 Bonn, Germany\\
$^{2}$Centre for Astrophysics and Supercomputing, Swinburne University of Technology, P.O. Box 218, Hawthorn, VIC 3122, Australia \\
}

\date{Accepted XXX. Received YYY; in original form ZZZ}

\pubyear{2020}

\begin{document}
\label{firstpage}
\pagerange{\pageref{firstpage}--\pageref{lastpage}}
\maketitle

\begin{abstract}
Machine learning methods are increasingly helping astronomers identify new radio pulsars. However, they require a large amount of labelled data, which is time consuming to produce and biased. Here we describe a Semi-Supervised Generative Adversarial Network (SGAN) which achieves better classification performance than the standard supervised algorithms using majority unlabelled datasets. We achieved an accuracy and mean F-Score of 94.9\% trained on only 100 labelled candidates and 5000 unlabelled candidates compared to our standard supervised baseline which scored at 81.1\% and 82.7\% respectively. Our final model trained on a much larger labelled dataset achieved an accuracy and mean F-score value of 99.2\% and a recall rate of 99.7 \%. This technique allows for high quality classification during the early stages of pulsar surveys on new instruments when limited labelled data is available. We open-source our work along with a new pulsar-candidate dataset produced from the High Time Resolution Universe - South Low Latitude Survey. This dataset has the largest number of pulsar detections of any public dataset and we hope it will be a valuable tool for benchmarking future machine learning models.
\end{abstract}


%%%%%%%%%%%%%%%%%%%%%%%%%%%%%%%%%%%%%%%%%%%%%%%%%%

%%%%%%%%%%%%%%%%% BODY OF PAPER %%%%%%%%%%%%%%%%%%

\section{Introduction}
Discovering a new pulsar can often lead to new and exciting science. Some examples include the discovery of PSR~B1257+12 \citep{1992Natur.355..145W}, which led to the discovery of the first set of extrasolar planets. The first binary pulsar PSR~B1913+16 \citep{1975ApJ...195L..51H} and the subsequent measurement of its orbital period decay provided the first indirect evidence of gravitational waves. The discovery of the first pulsar triple system (a pulsar orbiting two white-dwarfs) led to one of the most stringent test of the Strong Equivalence Principle (SEP), a prediction of general relativity  \citep{2020A&A...638A..24V}. More recently, PSR~J1141-6545 was used to infer Lense–Thirring precession (relativistic frame-dragging), a prediction of General Relativity \citep{2020Sci...367..577V}.
These examples are only some of the highlights that display the value of pulsar discoveries.
Therefore, in order to keep pushing the boundaries of fundamental physics, it is important that we continue the investigation of new techniques in order to enhance the discovery process.

Identifying radio pulsars involves finding usually broadband periodic signals in noise-dominated data. As pulsar signals pass through the interstellar medium (ISM) before arriving at radio telescopes, their radio emission is ``dispersed'' by the free electron content in the ISM. The amount of dispersion is proportional to the dispersion measure (DM) which is proportional to the integrated column density of free electrons between the pulsar and the observer. This creates a frequency-dependant delay such that lower frequency signals arrive later compared to higher frequency signals. Since the true DM of a pulsar is a \emph{priori} unknown, we typically de-disperse the data into multiple trial values. Once the data is dedispersed, periodic signals are identified in the timeseries by calculating a Fast Fourier Transform (FFT) \citep{CooleyTukey} or by using the Fast Folding Algorithm (FFA) \citep{1969IEEEP..57..724S}. Once the top candidates have been identified, these signals are folded at their respective spin-period and dispersion measure to form a pulsar candidate. Pulsar candidates are four-dimensional data-cubes consisting of time, frequency, rotational phase and power of a signal. These are the end products that are produced by most FFT based pulsar-search pipelines\footnote{Presto: \url{https://github.com/scottransom/presto} \\  Sigproc: \url{http://sigproc.sourceforge.net/}}. The final step is usually performed manually. Pulsar candidates are visualized as series of diagnostic plots (see Section \ref{subsection:features} for more details) which are used by pulsar astronomers to identify if a signal is from a genuine pulsar or not. 

Modern pulsar surveys, like the High Time Resolution Universe Low Latitude survey (HTRU-S Lowlat) \citet{2010MNRAS.409..619K} typically produce around 40 million pulsar candidates in one processing run. The increasing number of pulsar candidates can be attributed to multiple factors. Modern surveys tend to have higher time and frequency resolution. In addition, HTRU-S Lowlat also has a relatively long 72-minute integration time, which  leads to larger FFTs and the requirement of additional acceleration/template-bank trails in order to be sensitive to binary pulsars. We refer the readers to \citet{2016MNRAS.459.1104L} for a more in-depth review on this topic. Out of these 40 million candidates, only a few hundred thousand of them are expected to be real pulsar detections (multiple detections of known pulsars + new discoveries). This is because even after refining candidate lists to eliminate multiple occurrences of the same pulsar across many DM and acceleration/template bank trials and harmonics, many bright pulsars can appear in the sidelobes of many pointings and in a survey like HTRU-S Lowlat this results in a few hundred thousand detections of the known pulsars in the survey. Assuming an extremely optimistic average inspection time of one second per candidate, working 12 hours a day, it would take a human 2.5 years to go through the entire dataset. Future pulsar surveys using the Square Kilometre Array (SKA) telescope\footnote{\url{https://www.skatelescope.org/}}, are expected to increase this number further. Therefore, automated selection techniques that are optimised based on both speed and accuracy are of high importance for current and future pulsar surveys. Several papers have been written to address this topic. \citet{ralph_neural_net} used twelve hand-crafted numerical features/scores to describe each pulsar candidate. These twelve features were then attached to a multi-layer perceptron to identify pulsars in the Parkes multi-beam pulsar survey (PMPS) \citep{2001MNRAS.328...17M}. \citet{kj_lee_paper} introduced a candidate ranking scheme based on six quality factors that were selected based on domain knowledge. \citet{2014ApJ...781..117Z} developed a Pulsar Image Classification System (PICS) which is an ensemble machine learning model based on Convolutional Neural Networks (CNN), Support Vector Machines (SVM) and Artificial Neural Network (ANN). This technique was trained on candidates from the Pulsar
Arecibo L-band Feed Array (PALFA) survey \citep{2006ApJ...637..446C}, and were successfully applied for identifying pulsars in the Green Bank North Celestial Cap (GBNCC) survey \citep{2014ApJ...791...67S}. More recently, \citet{2019MNRAS.490.5424G}, used a combination of deep convolution generative adversarial network and support vector machines (DCGAN + L2SVM) to achieve excellent results for candidates in the HTRU Medlat and PMPS survey. However, all these techniques require a large number of labelled pulsar candidates in order to perform well. In practice, since the number of pulsar detections is only a small fraction of the total candidates, (<~1 per cent), previous works either under-sample the number of non-pulsars in their training data or over-sample the pulsar detections. In this paper, we present results from training a machine learning algorithm to address the practical scenario where we typically have a small amount of labelled data along with a large amount of unlabelled data. This is called Semi-Supervised learning. Past applications using a similar approach in astronomy include applying Semi-Supervised learning on data from the Very Long Baseline Array (VLBA) Fast Radio Transients Experiment (V-FASTR) for radio pulsar candidate classification \citep{jones2012big, bue2014astronomical}, applying a Semi-Supervised distributed algorithm called Co-Training, Distributed, Random Incremental Forest (CoDRIFt)\footnote{\url{https://github.com/tdevine1/cluster/tree/master/code/CoDRIFt}} for single pulse pulsar candidate classification \citep{devine2020searching} and another study which uses a Semi-Supervised Deep Convolutional Neural network to classify radio galaxy images \citep{ma2019classification}. We also compare our results to the purely supervised approach as done by previous works.

\section{Methods}
\subsection{Machine Learning}
Machine learning is a branch of computer science that deals with solving problems by learning through experience. In the classical setup, a human defines all the steps necessary for a computer to solve the problem. However, for complex tasks when it is not trivial to come up with a model to map the input data to our desired output, it is often desirable to learn from the data itself. This process of learning through experience is usually called ``training'' an algorithm. 

There are broadly three classes of machine learning that are relevant for the work in this paper. 
\begin{enumerate}
    \item Supervised Learning: In supervised learning, we have data and its corresponding label, which in our case is a binary label between pulsar and non-pulsar signals. To the best of our knowledge, all the currently published papers in pulsar candidate classification fall under this category.
    \item Semi-Supervised Learning: Semi-supervised learning is a branch of machine learning that combines a small amount of labelled data along with a large number of unlabelled data in order to obtain better learning performance. This is the problem we are trying to tackle in this paper.
    \item Unsupervised Learning: In unsupervised learning, no labels are provided during training. It is up to the algorithm to find useful structure in the input data. 
\end{enumerate}
\subsection{Artificial Neural Network (ANN)}
ANNs are a class of supervised machine-learning algorithms that are commonly used for classification tasks. Variants of this network have been used previously in solving the pulsar candidate classification problem in \citet{ralph_neural_net, 2012MNRAS.427.1052B, 2014ApJ...781..117Z, 2018A&C....23...15B}. This algorithm has also been used in this paper for comparing our proposed architecture to the standard supervised learning case. We briefly summarise the different components of an ANN and its operation. For a more thorough explanation, refer to chapter five of \citet{10.5555/1162264}.  

\begin{figure}
\includegraphics[width=\columnwidth]{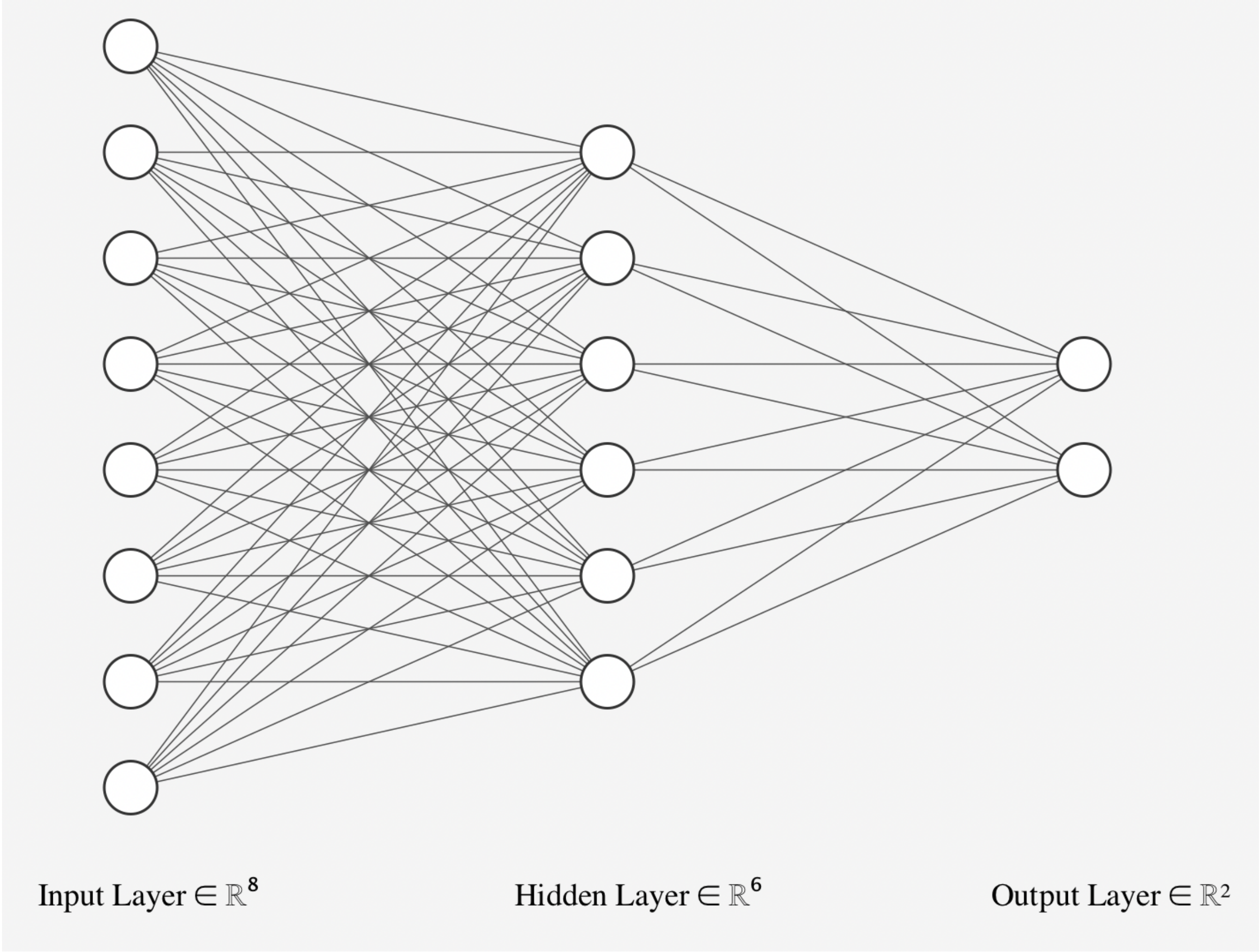}
    \caption{Multi-layer Artificial Neural Network with 8 input units, 6 hidden units and 2 output units. In shorthand, they are usually written as a 8:6:2 network. This diagram was created using an open source tool called NN-SVG\protect\footnotemark \citep{LeNail2019}.}
    \label{ann_diagram}
\end{figure}

The simplest unit of an ANN is a neuron. These neurons are loosely inspired by biological neurons in the sense that there are input(s) to the neuron, an activation function and an output. Neurons are usually grouped together in layers. The first layer (often called the input layer) of the ANN is usually attached to the image or data we are interested to predict on and the last layer (often called output layer) is usually attached to the label we want to predict. Figure \ref{ann_diagram} is an example of single layer neural network, where there is  one input layer, hidden layer and output layer respectively. Each neuron of a layer is connected to all the neurons of the next layer. A neural network with several hidden layers is usually referred to as a deep neural network or multi-layer perceptron (MLP). Assume we have an input vector $\vec{X}$ of 8 elements $\{X_1, X_2, ... X_{8}\}$ which is passed onto a neuron in the next layer. This neuron then calculates a weighted sum of all the values of $\vec{X}$ and applies a non-linear activation function on it. Mathematically, this output $\hat{y}$ can be written as:
\begin{equation}
    \hat{y} = g(w_0 + \Sigma_{i=1}^m X_i w_i), \mathrm{where,}
\end{equation}
w$_i$ is the weight of the i$\mathrm{^{th}}$ neuron which decides the relative importance of a neuron, w$_0$ is the bias term which is a trainable constant value for each layer, $g$ is the activation function used. The purpose of an activation function is to decide if the neuron should be activated or not. This function helps in normalising the weighted sum values and additionally they introduce non-linearities to the network. Some of the activation functions used in this paper include the sigmoid function $\left(f(x) = \frac{1}{1 + e^{-x}}\right)$, tanh $\left(f(x) = \frac{e^{-2x} - 1}{e^{-2x} + 1}\right)$ and Rectified Linear Unit 'ReLU' $\left(f(x) = max(0,x)\right)$. \\
\\
The process by which a neural network ``learns'' is by minimising a loss function. Loss functions are defined as the difference between the predicted output of a neural network to the ground truth labels. Since, we are dealing with a binary classification problem, the output of our neural network is a probability value between 0 and 1 for a candidate to be a pulsar. A value closer to either extremum indicates high confidence in our prediction. Our goal is to minimise the cross-entropy loss between our predicted labels and true labels. We use the standard softmax function $\left( \sigma(\vec{z})_i = \frac{e^{z_i}}{\sum_{j=1}^K e^{z_j}} \right)$ for converting values into probabilities. Here, $\vec{z}$ is the input vector passed on-to the softmax function and K is the number of classes in the classifer. In practice, these loss functions are minimised in an iterative fashion by calculating their negative gradients and propagating it backwards to the network using a process called back-propagation \citep{1986Natur.323..533R}.
\begin{comment}

\begin{table}
	\centering
	\caption{Activation functions used in this paper.}
	\label{tab:activation functions}
	\begin{tabular}{ll} 
		\hline
		Activation Function & f(x)  \\
		\hline
		\vspace{0.2cm}
		Sigmoid & $\frac{1}{1 + e^{-x}}$\\
		\vspace{0.2cm}
		
		tanh &  $\frac{e^{-2x} - 1}{e^{-2x} + 1}$ \\
		\vspace{0.2cm}
		Rectified Linear Unit (ReLU) & max(0,x) \\
		%3 & 5 & 7 & 9\\
		\hline
	\end{tabular}
\end{table}
\end{comment}

\footnotetext{\url{https://alexlenail.me/NN-SVG/index.html}}
\subsection{Convolutional Neural Network (CNN)}
\begin{figure}
	\includegraphics[width=\columnwidth]{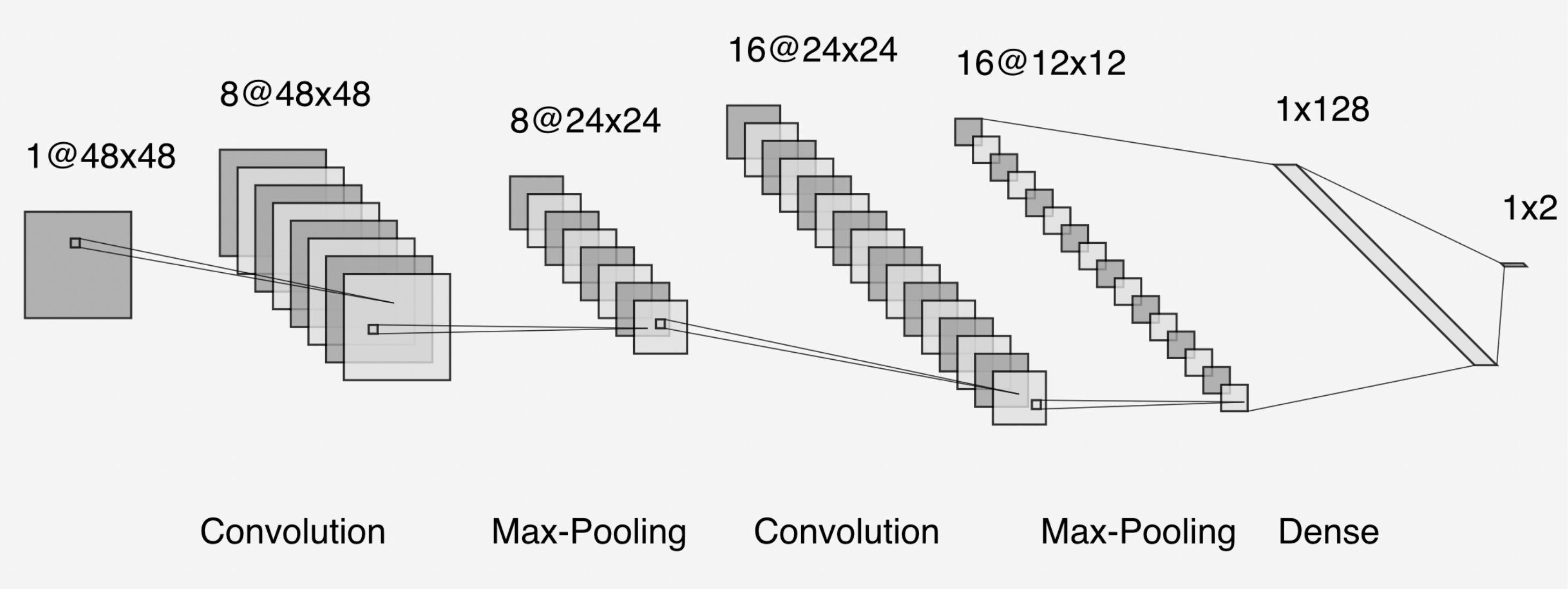}
    \caption{Schematic diagram of a typical CNN. Here the input image is a grey scale image of size 48x48 and the output contains two nodes for the binary class labels. 8 and 16 filters of size 4x4 were used for the convolutional layers. These act as feature extractors that identify import patterns in the input image. This diagram was created using an open source tool called NN-SVG \protect\footnotemark \citep{LeNail2019}.}
    \label{cnn_diagram}
\end{figure}

CNNs are a class of machine learning algorithms that are commonly applied in the field of image classification. One of the earliest applications of CNNs was the LeNet-5 network, which was successfully used to recognise hand-written digits \citep{lecun-98}. CNNs have also been successfully applied to the pulsar candidate classification problem previously in \citet{2014ApJ...781..117Z, 2019MNRAS.490.5424G}. An example of a CNN is given in Figure \ref{cnn_diagram}. The major difference here is that the fully connected layers have been replaced with convolutional layers. These convolutional filters act as feature extractors to identify important parts of the input image. The convolutional layer is typically followed by a max-pooling operation, where we find the maximum value of preceding neuron cluster and store them to a single neuron in the current layer. For example, if we have a 48x48x1 tensor, after a 2D max pooling operation of 2x2, the tensor's size changes to 24x24x1. This is done to constrain the dimensionality of the network while propagating only important information to the next layer. This is usually followed by a activation function, typically ReLU. Many such convolution, max-pooling and activation function layers can be concatenated together to a form a deep CNN. This is usually followed by a fully connected layer also known as a dense layer which is then finally connected to the output layer. The output layer for a classification problem is the amount of class labels we have in our data. Additionally for deep CNNs, a dropout layer is typically added after the max-pooling operation, which randomly drops off a certain percentage of the preceding nodes. This is done so that the neural network learns to better generalize its performance across the entire data. This is used as a regularization technique to avoid overfitting. CNNs are also trained using back-propagation. However, in practice since it is computationally difficult to calculate the gradient of the loss function for all images in the training data, we typically divide the data into mini-batches and use the stochastic gradient descent algorithm.

\footnotetext{See footnote 4.}

\subsection{Generative adversarial network (GAN)}
\label{gan_explained}
GANs are a class of machine learning algorithms \citep{NIPS2014_5423} in which two neural networks are trained simultaneously with opposing goals. They act against each other as adversaries in a minmax two-player game. A generative model \emph{G} is tasked with generating new data that captures the distribution of the input data. A discriminator model \emph{D} is tasked with classifying samples as either REAL (that belong to the original data distribution) or FAKE (samples generated by \emph{G}). Assuming $V (\emph{D}, \emph{G})$ is the value/loss function we are interested in then the first term in Equation \ref{gan_equation}, is the expectation of the logarithm of \emph{D}'s predictions when an input is from the real data sample. \emph{D}'s goal is to maximise this term. The second term represents one minus the expectation of the logarithm of \emph{D}'s predictions when data generated by \emph{G} is passed onto to \emph{D}. The goal of \emph{D} is to maximise the second term but the goal of \emph{G} is to minimise this term. This sets up the adversarial framework. In the ideal case, the generator perfectly samples the distribution of the input data distribution, and the discriminator output equals to 1/2.

\begin{equation}
\label{gan_equation}
\begin{aligned}
    \underset{G}{min}\underset{D}{max} V(D, G) = \mathbb E_{x \sim p_{data (x)}}[log \emph{D (x)}]  \\ + \mathbb E_{z \sim p_z(z)}[log (1-D(G(z)))]
\end{aligned}
\end{equation}

In practice however, we use a minibatch stochastic gradient descent method and train the generator and the discriminator alternatively. The algorithm and the proof for the convergence of the algorithm can be found in \citet{NIPS2014_5423}, however for the benefit of the reader we briefly summarise the training of this network in algorithm \ref{alg:gradient descent}.\\

\begin{algorithm}
\label{alg:gradient descent}
\SetAlgoLined
\DontPrintSemicolon
\SetAlgoRefName{1}
\setstretch{1.25}
%\SetAlgoSkip{smallskip}]
\caption{Minibatch stochastic gradient descent algorithm used for training a GAN.}

\For{\textsf{\upshape N training epochs}}{
\For{\textsf{\upshape k batches}}{

Fix the weights of \emph{G} and update \emph{D} \;
%\\
Sample a mini-batch of m noise samples $\left\{z^{(1)}, ..., z^{m}\right\}$ from a noise distribution p$_z(z)$\;

Sample a mini-batch of m real data samples $\left\{x^{(1)}, ..., x^{m}\right\}$ from the data distribution~p$_{data(x)}$\;

Update the Discriminator by ascending its stochastic gradient. 
\begin{equation*}
    \nabla_{\theta_d} \frac{1}{m}\Sigma_{i=1}^m \left[log D (x^{(i)} + log (1 - D(G(z^{(i)}))) \right].
\end{equation*} 
%\\
Fix the weights of \emph{D} and update \emph{G} \;
Sample a mini-batch of m noise samples $\left\{z^{(1)}, ..., z^{m}\right\}$ from a noise distribution p$_z(z)$. \;
Update the Generator by descending it's stochastic gradient.
\begin{equation*}
    \nabla_{\theta_g} \frac{1}{m}\Sigma_{i=1}^m \left[ log (1 - D(G(z^{(i)}))) \right].
\end{equation*}}}
\end{algorithm}

\begin{figure*}
	\includegraphics[width=1.5\columnwidth]{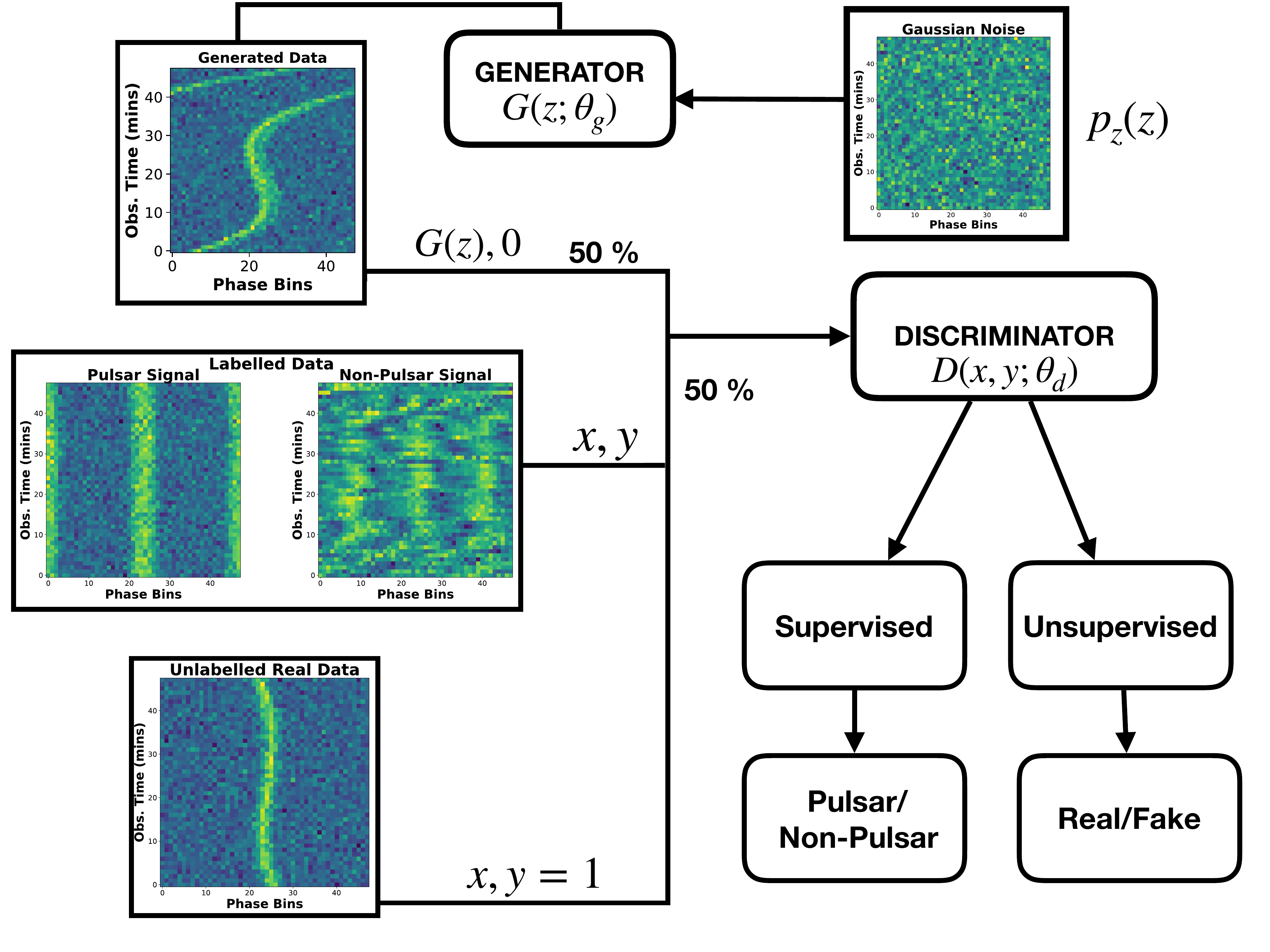}
    \caption{Schematic of the SGAN Architecture used in this paper. The generator is initialised with a noise (a.k.a latent vector) variable, which it then transforms to a fake generated image. The discriminator is fed images from three sources i) Labelled Pulsar candidates, ii) Unlabelled pulsar candidates provided with a positive label and iii) Fake generated images from the generator provided with a negative label. The discriminator is  tasked with minimising both the supervised and unsupervised loss function.}
    \label{sgan_figure}
\end{figure*}

GANs have been used successfully in a wide range of tasks ranging from computer vision, where they have been used for generating photo-realistic images \citep{2018arXiv181204948K}, converting text to images \citep{Reed2016GenerativeAT}, used as feature extractors for unsupervised learning \citep{2015arXiv151106434R}. In astronomy, GANs have been successfully demonstrated to recover features from astrophysical galaxy images beyond the deconvolution limit \citep{2017MNRAS.467L.110S}, create high-fidelity weak lensing convergence maps \citep{2019ComAC...6....1M}, modeling exoplanet atmospheres \citep{2018AJ....156..268Z} and more recently also in pulsar candidate identification \citep{2019MNRAS.490.5424G}. The standard framework of GAN described here usually comes under the category of unsupervised learning as no class labels are provided while training the network. 
\subsection{Semi-Supervised
Generative Adversarial Network (SGAN)}
\label{section:sgan}
SGANs are a variant of GAN where we can leverage both the ability of the generator to create realistic samples and the readily available unlabelled pulsar candidates to solve the semi-supervised classification problem. 
In the standard GAN problem, the output of \emph{D} is a probability for the input image belonging to the training set. We modify this standard architecture slightly by adding the samples from \emph{G} into our training set. We label them as a new generated class say K+1 where K is the total number of classes in our original classification problem. We then change the dimension of \emph{D}'s output from a binary classification output to a multi-class classification output \{Pulsar, Non-Pulsar, Fake Data\}. The main advantage of this technique is that we can now learn from our pulsar survey's unlabelled data.

There are three major components of this network, a supervised discriminator, an unsupervised discriminator and an unsupervised generator. The setup for the unsupervised discriminator and generator are similar to the standard GAN architecture discussed in Section \ref{gan_explained}. The supervised discriminator is provided with class labels (Pulsar or Non-Pulsar) that are available from our training set. The remaining unlabelled pulsar candidates were provided to the unsupervised discriminator with a positive label (`1') and generated fake candidates from \textit{
G} were provided with a negative label (`0'). For every training epoch, 50 per cent of the samples were taken from the generator and 50 per cent of the samples were taken from a combination of both labelled and unlabelled candidates. A schematic of this architecture can be found in Figure \ref{sgan_figure}.

Mathematically, the loss function for SGANs can be written as

\begin{equation}
\begin{split}
        \mathrm{L} &= - \mathbb E_{x,y\sim p_{data(x,y)}}[log \, \mathrm{p_{model}}(y|x)] \\
        &- \mathbb E_{x \sim G}[\mathrm{log \, p_{model}} (y = K + 1 | x)] \\
        &= \mathrm{L_{supervised} + L_{unsupervised}, where} \\
        \mathrm{L_{supervised}}&= -\mathbb E_{x,y \sim p_{data}(x,y)} log \, \mathrm{p_{model}}(y|x,y < K+1) \\
        \mathrm{L_{unsupervised}}&= -\mathbb E_{x \sim p_{data}(x)} log \, [1 - \mathrm{p_{model}}(y=K+1|x)] \\ 
        &+ \mathbb E_{x \sim G} log \, [\mathrm{p_{model}}(y = K+1|x)].
    \end{split}
\end{equation}
The first term in the loss function is the so called supervised loss, $\mathrm{L_{supervised}}$. This is similar to the standard loss function of any supervised classification model ($\mathrm{p_{model}}$) where x is the data and y is it's corresponding label. The unsupervised loss term ($\mathrm{L_{unsupervised}}$) consists of two parts, the first part corresponds to one minus the expectation that the model will output the new `FAKE' class (K+1) given that the data is real and the second term is the expectation that the model will correctly identify the newly generated `FAKE' class given that that the data comes from the generator. If we substitute, $\emph{D(x)} = 1 - \mathrm{p_{model}}(y=K+1|x)$ into the unsupervised loss term, we notice that this is equivalent to the regular GAN's loss function described in Equation \ref{gan_equation}. The goal of SGANs is to minimise these two loss functions jointly. The key to the unsupervised term being useful is that the generator needs to be trained to approximate the input data distribution which in turn minimises the first term of the unsupervised loss function. The formalism for SGANs described here and several practical implementation tricks we used, were largely inspired by the work of \citet{NIPS2016_6125}.

\subsection{Data Preprocessing and Features Used}
\label{subsection:features}
A pulsar candidate is a four dimensional data cube of frequency channels, time, power and rotational phase of the signal. Since it is inconvenient to visualize four dimensional data, the convention is to plot various two-dimensional and one-dimensional projections of this data-cube to decide if a signal is really from a pulsar or not. The four feature plots pulsar astronomers most often use are:
\begin{enumerate}
    \item Pulse Profile: This one dimensional intensity curve is created by integration over both time and frequency axes while preserving phase. Most real pulsars tend to have a one or multiple narrow peaks. However, there are some known exceptions. Some known pulsars, especially millisecond pulsars (MSPs) tend to have broader or close to sinusoidal profiles.
    \item Frequency-phase Plot: This two dimensional plot is created by integrating over the time axis only. Real pulsars tend to be broadband, therefore, we expect a persistent bright signal (vertical line) across all sub-bands. However, pulsar scintillation caused by the interstellar medium can sometimes increase or decrease the signal in some frequency channels (e.g. PSR~B0355+54 \citet{2018MNRAS.476.5579X}).
    
    \item Time-Phase Plot: This two dimensional plot is created by integrating across the frequency axis only. We expect most pulsars to be persistent across observing time. There are some notable exceptions, for example a nulling pulsar like PSR~J1727-2739 \citep{2016A&A...592A.127W}, relativistic binary pulsars which can have quadratic or cubic residuals in the time-phase plot or mildly accelerated pulsars  where the acceleration falls between trial values.
    \item DM-Curve: This is a one dimensional plot to find the best-fit dispersion measure value. In order to produce this, the candidate data is dedispersed around a few trial DM values from the DM used to fold the candidate. For each trial, it calculates the chi-squared of the dedispersed pulse profile against a horizontal line fit. A large chi-squared value is an indication that the signal deviates from white noise. Since pulsars are non-terrestrial signals, we expect the signal to peak at a non-zero DM value. The sharpness of the DM curve depends on the duty cycle of the pulsar. 
    
\begin{comment}
constant power vs time.
\end{comment}    
\end{enumerate}

\begin{figure*}
    \centering
    \subfigure[]{\includegraphics[width=\columnwidth]{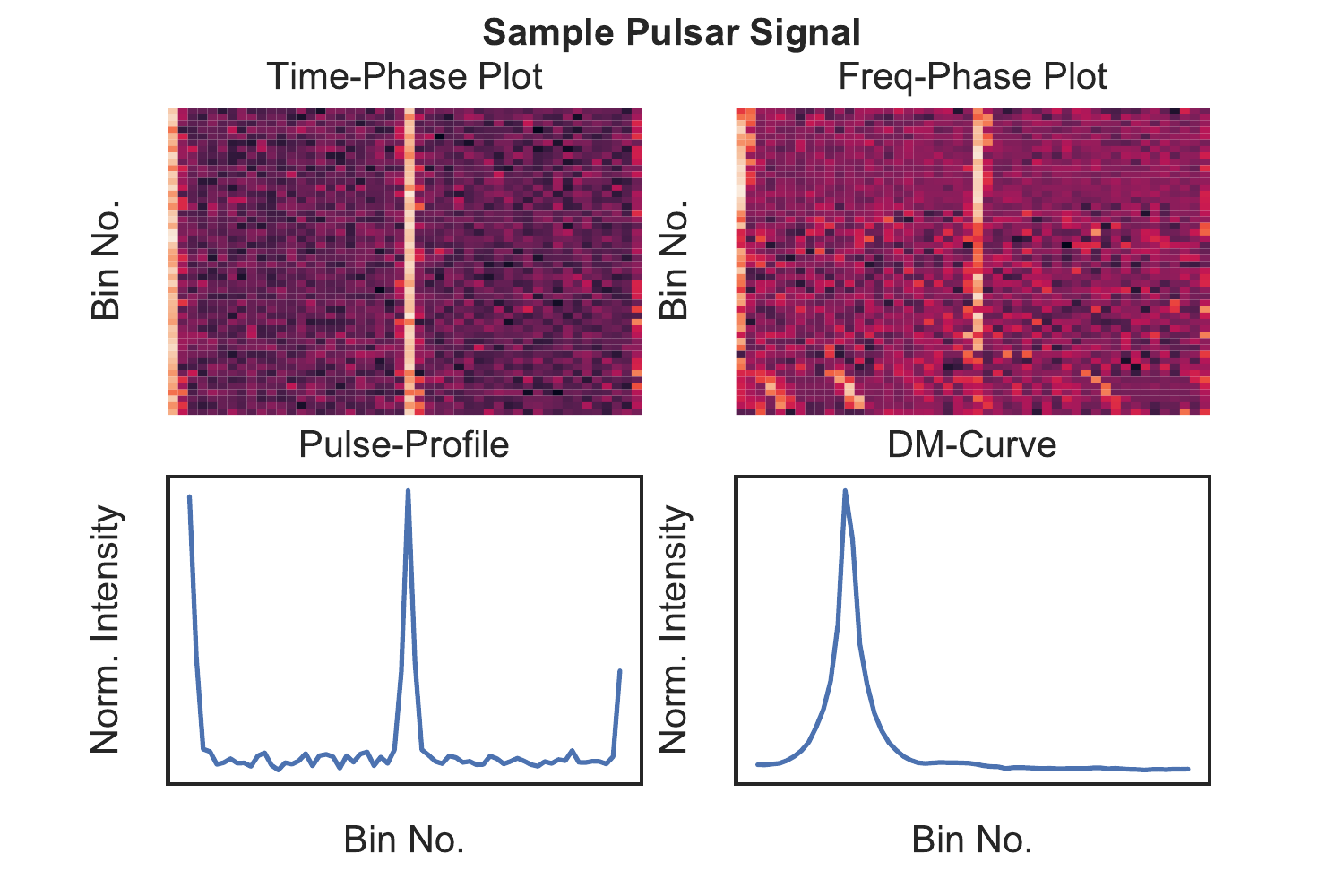}} 
    \subfigure[]{\includegraphics[width=\columnwidth]{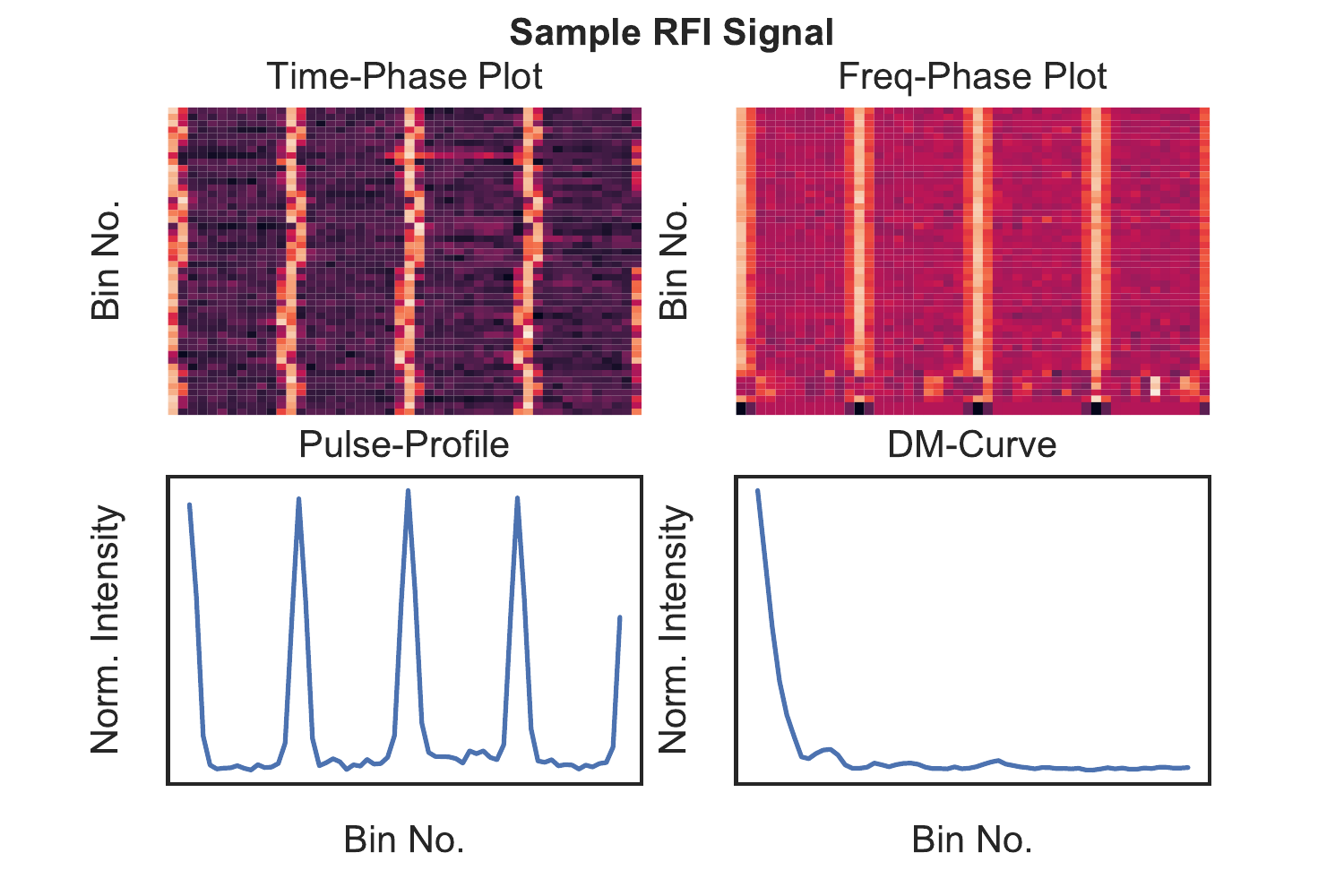}}
    \subfigure[]{\includegraphics[width=\columnwidth]{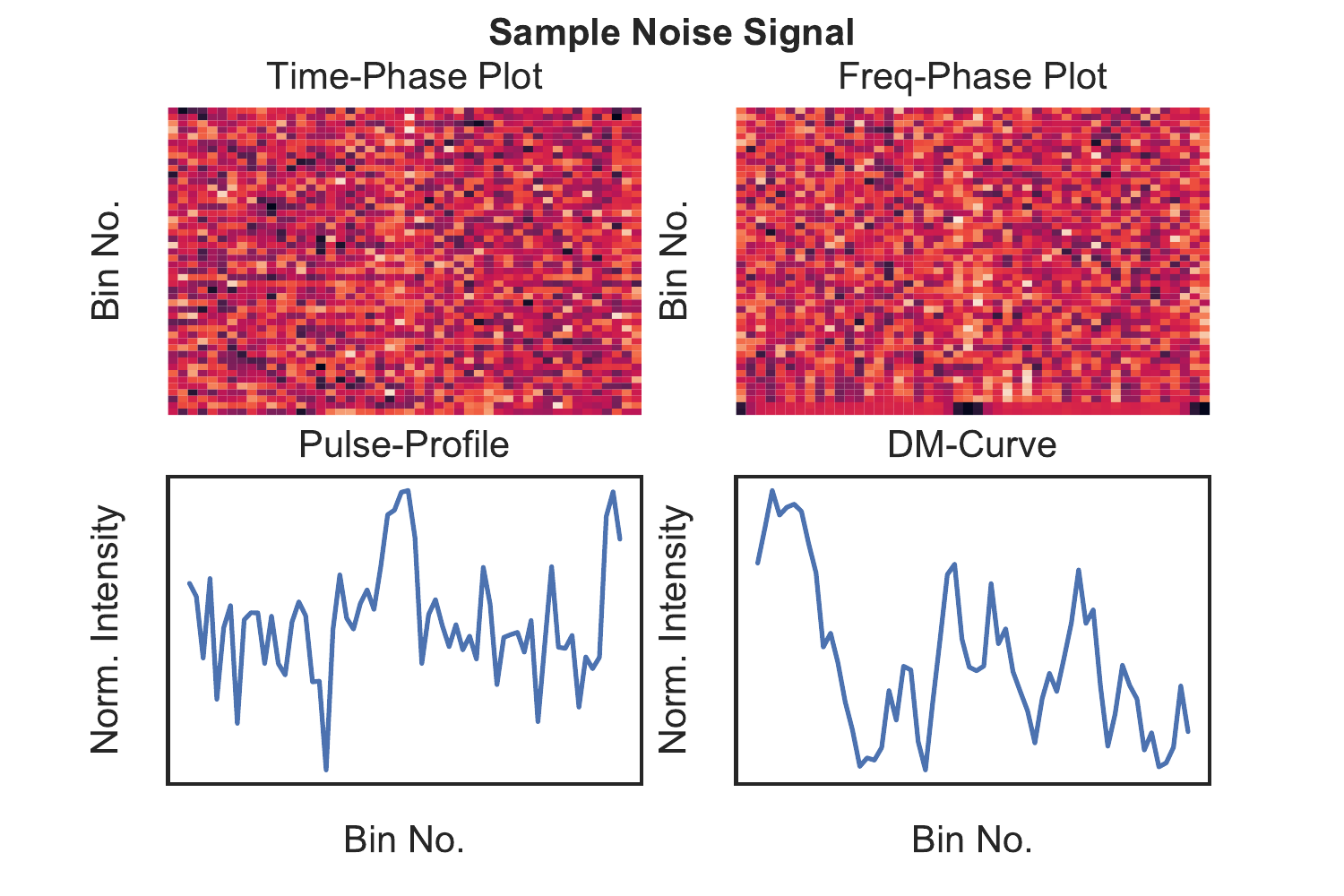}}
    \caption{Sample data of the different types of signal in our training data: (a) Signal from a known pulsar, (b) Signal from a broadband RFI signal and (c) White Noise signal. Under each class label, we see a plot from the four features that was used to train all the machine learning algorithms used in this paper.}
    \label{sample_data}
\end{figure*}

We used the four features mentioned above to train the semi-supervised network. Before the data is passed onto the network, it is important to standardise the data, so that the algorithm is agnostic to spin-period, dispersion measure, observing frequency and integration time of an observation. We use the publicly available data pre-processing code made available by \citet{2014ApJ...781..117Z} for our work.\footnote{\url{https://github.com/zhuww/ubc_AI}} 
%We used the same four features and data standardisation %techniques as described in \citet{2014ApJ...781..117Z}, %while training our semi-supervised network. 
An example of the four features and the different types of signal in the training set is shown in figure \ref{sample_data}. 
In order, to have the same number of bins for all candidates, the code down-samples and interpolates the data using linear interpolation for the 1-D plots and spline interpolation for the 2-D plots. The data is also normalized to have zero median and unit variance. We use 60 bins for the DM-curve and 64 bins for the pulse-profile. The time-phase and frequency-phase plots were resampled to a size of 48x48 bins. The bin sizes for the different features were chosen to maintain consistency with \citep{2014ApJ...781..117Z} for our eventual comparisons.

\subsection{Metrics Used}
\label{sec:metrics used}
We use a combination of seven metrics to evaluate our different machine learning models. Since we have reformulated our network to a binary classification problem, there are four possible scenarios when we compare the predicted labels to the true labels. An example of this is shown in table \ref{tab: confusion_matrix_explained}. This is usually referred to as confusion matrix in the literature. Based on this we calculate the following metrics:

%\makegapedcells
\begin{table}
\makegapedcells
	\centering
	\caption{Confusion Matrix for the Pulsar Candidate Classification problem.}
\begin{tabular}{cc|cc}
\multicolumn{2}{c}{}
            &   \multicolumn{2}{c}{Predicted Label} \\
    &       &   Pulsar &   Non-Pulsar              \\ 
    \cline{2-4}
\multirow{2}{*}{\rotatebox[origin=c]{90}{True Label}}
    & Pulsar   & True Positive (TP) & False Positive (FP)                 \\
    & Non-Pulsar    & False Negative (FN) & True Negative (TN)                \\ 
    \cline{2-4}
   
    \end{tabular}
    \label{tab: confusion_matrix_explained}
 \end{table}

\begin{comment}
\begin{table}
	\centering
	\caption{Confusion Matrix}
	\label{tab:confusion_matrix}
	\begin{tabular}{lcc} % four columns, alignment for each
		\hline
		True Labels & Pulsar & Non-Pulsar \\
		Predicted Labels & & \\
		\hline
		Pulsar & True Positive (TP) & False Positive (FP) \\
		Non-Pulsar & False Negative (FN) & True Negative (TN) \\
		%3 & 5 & 7 & 9\\
		\hline
	\end{tabular}
\end{table}
\end{comment}
\begin{enumerate}
    \item The simplest metric to calculate is accuracy. While this can be a useful metric to evaluate our model, care must be taken to ensure that our training data is balanced. In unbalanced training datasets, a high accuracy score alone is not an indication of a useful machine learning model. \\ 
    
    Accuracy = $\mathrm{\frac{Number \, of \, Correct \, Predictions}{Total \, Predictions}}$
    = $\mathrm{\frac{TP \, + \, TN}{TP \, + \, TN \, + \, FP \, + \, FN}}$
    \\
    \item Precision is defined as the fraction of true positives among all the positive label outputs of the model. This is a useful metric if our goal is to minimise the number of false positive (RFI + white noise) candidates which often translates to less human hours spent in candidate inspection. \\
    
    Precision = $\mathrm{\frac{TP}{TP \, + \, FP}}$
\\
    \item Recall or Sensitivity is the ratio of true positives to the total sum of true positives and false negatives. A high recall rate indicates that our model was successful in extracting most of the pulsar signals from our data. \\
    
    Recall = $\mathrm{\frac{TP}{TP \, + \, FN}}$ \\
    \\
    \item F-score or F$_1$ score is defined as the harmonic mean of precision and recall. \\
    
     F-score = $\mathrm{2 \,   \frac{Precision \, \cdot \, Recall}{Precision \, + \, Recall}}$
    \\
    \item False positive rate (FPR) is analogous to precision. It is defined as the ratio of false positives to the total sum of false positives and true negatives. Unlike the other metrics used in this paper, a lower score of FPR is more desirable. \\
    
    FPR = $\mathrm{\frac{FP}{FP \, + \, TN}}$
    \\
    \item Specificity is defined as the ratio of true negatives to the total sum of true negatives and false positives. This is analogous to the recall rate we defined earlier. A high specificity rate indicates that our model was successful in extracting most of the non-pulsar signals from our data. \\ 
    
    Specificity = $\mathrm{\frac{TN}{TN \, + FP}}$
    \\
    \item G-Mean is defined as the geometric mean of recall and specificity. \\
    
    G-Mean = $\mathrm{\sqrt{2 \, \cdot \, Recall \, \cdot \, Specificity}}$

\end{enumerate}

\begin{comment}

\begin{table}
	\centering
	\caption{Detailed Comparison Between CNN and SGAN}
	\label{tab:sgan_versus_cnn}
	\begin{tabular}{lcc} % four columns, alignment for each
		\hline
		Architecture & CNN & \textbf{SGAN}\\
		\hline
		Accuracy & 0.763 & 6 \\
		Precision & 0.809 & 3 \\
		Recall & 0.809 & 3 \\
		
		F-score & 0.809 & 3 \\
		
		Specificity & 0.809 & 3 \\
		
		G-Mean & 0.809 & 3 \\
		
		FPR & 0.809 & 3 \\
		%3 & 5 & 7 & 9\\
		\hline
	\end{tabular}
\end{table}

\end{comment}



\begin{figure*}
	\includegraphics[width=0.85\linewidth]{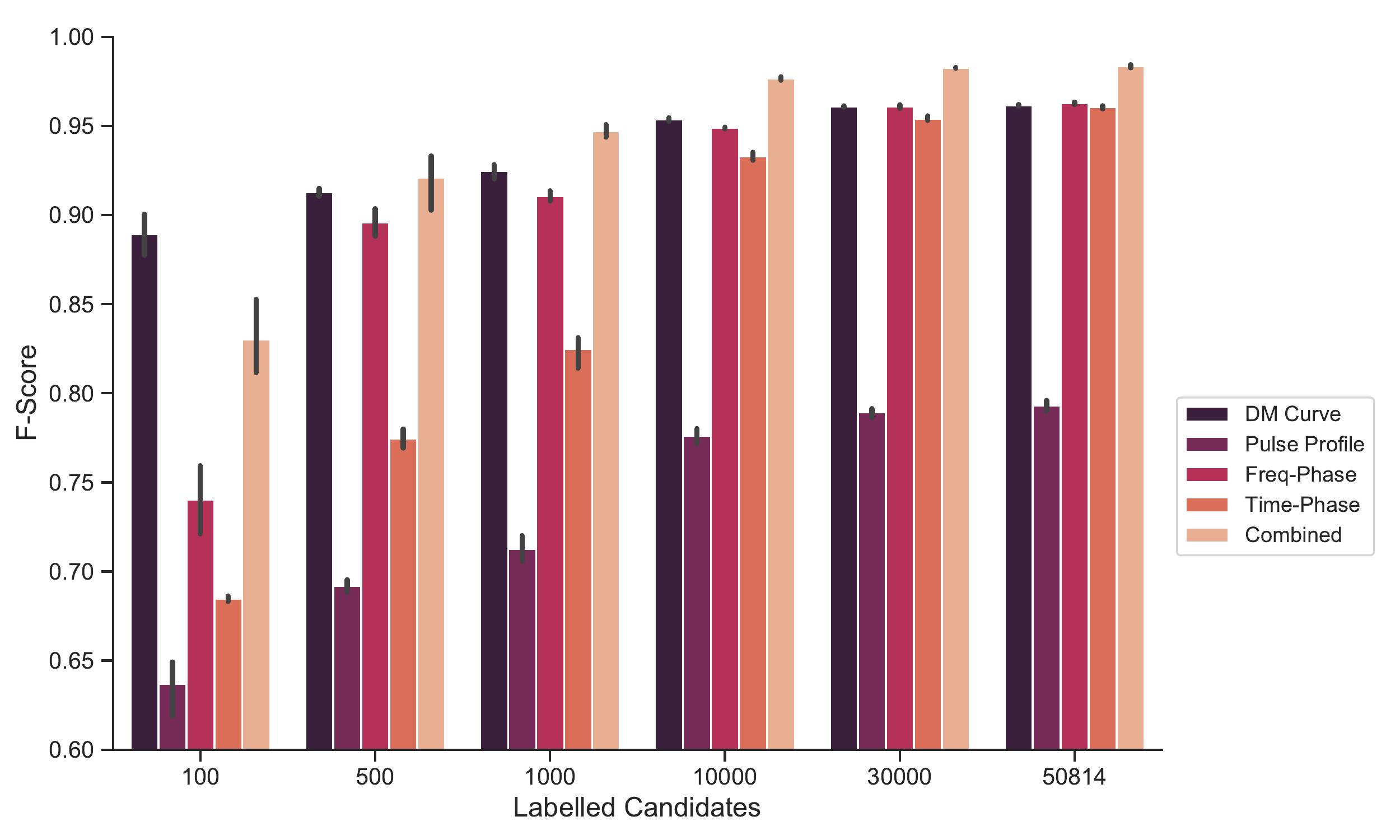}

\caption{Mean F-score values of our supervised baseline model for varying number of labelled candidates. We find that ``DM-Curve'' is the best performing feature for labelled candidates below 500 and ``Freq-Phase'' as the best performing feature for larger number of labelled candidates. The combined model is a logistic regression model fit on the training data. Each model was averaged across five different batches of labelled data. The vertical bar displays the 95\% confidence interval of our results.}
\label{fig:sup_performance}
\end{figure*}

\section{Data Used in this study}
We used observations from the High Time Resolution Universe South Low-Latitude Survey (HTRU-S Lowlat) to generate pulsar candidates. HTRU-S Lowlat is one part of the entire HTRU Survey that focuses on the inner galactic plane covering galactic longitude \ang{-80} < \emph{l} < \ang{30} and galactic latitude |\emph{b}| < \ang{3.5}. The observations were recorded for an integration time of 72 minutes with a frequency bandwidth of 400~MHz using the 64-m Parkes Radio Telescope. We refer the reader to \citet{2010MNRAS.409..619K} for a full list of the observational setup and system configuration. To date, HTRU-S Lowlat has discovered >100 new pulsars. A full list of the initial discoveries and timing solution for these pulsars can be found in \citep{2015MNRAS.450.2922N, 2020MNRAS.493.1063C}. The pulsar candidates used for training all our models were generated from the re-processing of the HTRU-S Lowlat survey using the stochastic template-bank algorithm \citet{2009PhRvD..80j4014H}, and folded using the PRESTO software suite \citet{2011ascl.soft07017R}.  The aim of the re-processing pipeline is to find compact relativistic binary pulsars that may have been missed by the first pass time-domain segmented acceleration search pipeline \citet{2015MNRAS.450.2922N}. The total number of pulsar candidates produced by the re-processing pipeline for the entire survey is around 40 million. We selected 84,691 candidates that were labelled by eye to have approximately similar number of pulsar and non-pulsar candidates. We carefully chose pulsar candidates of different significance levels in order to create a diverse labelled candidate dataset. Our lowest detection significance of a true pulsar candidate is 4.3 sigma. The breakdown of candidates have been shown in Table \ref{tab:train and test dataset}. To the best of our knowledge, this labelled dataset has the largest number of pulsar detections out of all the publicly available pulsar candidate datasets. Labelled Pulsar candidates are critical for training machine learning algorithms as they are scarce (<~1 per cent) compared to the total candidates produced in a pulsar survey.

\begin{table}
	\centering
	\caption{Total Number of pulsar and non-pulsar candidates in our labelled dataset. Candidates marked as `Pulsar' also include harmonics and multiple detections of the same pulsar.}
	\label{tab:train and test dataset}
	\begin{tabular}{lccr} % four columns, alignment for each
		\hline
		 & Pulsar & Non-Pulsar & Total \\
		\hline
		Training Set & 25,888 & 24,926 & 50,814 \\
		Validation Set & 6,497 & 6,207 & 12,704  \\
		Testing Set & 10,676 & 10,497 & 21,173 \\ 

		\hline
	\end{tabular}
\end{table}

\begin{figure}
\begin{center}
	\includegraphics[width=0.45\textwidth]{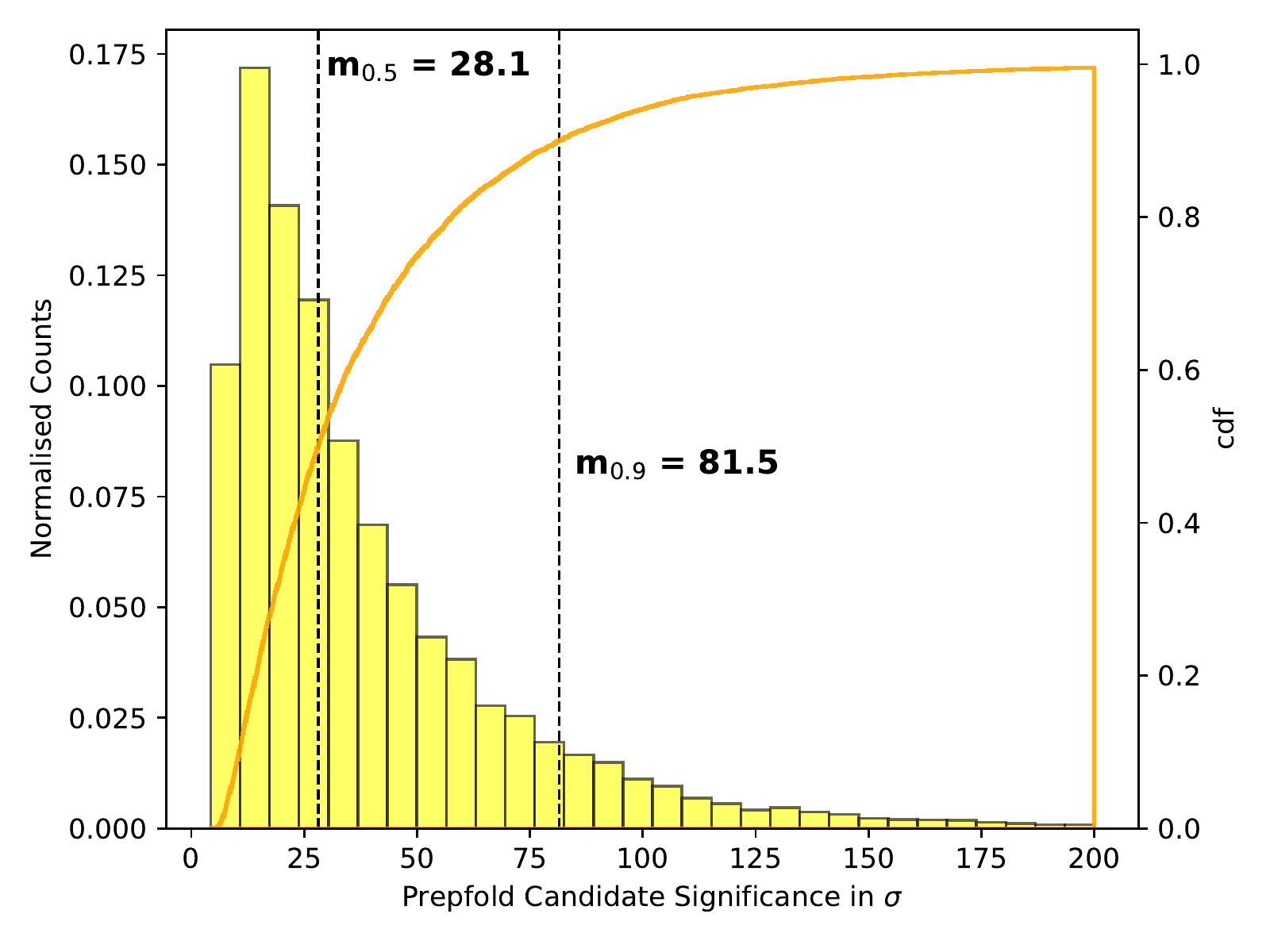}
	\caption{Histogram of the detection significance levels in sigma (i.e equivalent gaussian significance) of pulsars in our test dataset. The orange line shows the cumulative distribution function of the candidate significance levels. Vertical dashed lines indicate the median and 90th percentile values of the pulsar detection significance. For plotting purposes, we deliberately clipped out outlier candidates with a sigma greater than 200.}
	\label{fig:hist_sigma}
\end{center}
\end{figure}

\section{Results}
We start by splitting our entire labelled dataset into a train, validation and test dataset (60\% train, 15\% validation and 25\% test, see table \ref{tab:train and test dataset}). The test dataset was never seen by the network while training. This dataset is only used in the end as a benchmark to evaluate all the different experimental setups described below. See figure \ref{fig:hist_sigma}, for the detection significance levels of pulsars in our test set. The validation dataset was used to tune hyperparameters of the different architectures and to select the best model during training. We train all the models separately on each of the four features described in section \ref{subsection:features}. Our software was built using Keras\footnote{\url{https://keras.io/}} \citep{chollet2015keras}, a high-level open-source neural network library with Tensorflow 2.0 backend \footnote{\url{https://www.tensorflow.org/}} \citep{tensorflow2015-whitepaper}. 
\subsection{Supervised Learning Baseline}
Our first goal was to build a model based on supervised learning that would act as our comparative baseline. For this, we trained a convolutional neural network (CNN) on the ``Time-Phase'' and ``Freq-Phase'' features and a multi-layer perceptron (MLP) for the ``DM-Curve'' and ``Pulse Profile'' features. Each network was trained for a total of 1000 epochs for different amounts of labelled data, saving only the model that produced the highest accuracy on our validation dataset. For each batch of labelled candidates, we split the training data to have an equal number of pulsar and non-pulsar signals. For example, 100 labelled candidates implies that the training data had 50 pulsar and 50 non-pulsar candidates. Since, the results are dependent on the subset of training samples used while training, we randomly selected five different combinations of labelled candidates and report the average values. The mean F-score performance of each of the four features is shown in Figure \ref{fig:sup_performance}.  We observe that in the regime of extremely limited labelled data (labelled candidates <= 500), ``DM Curve'' acts as the best discriminator between pulsar and non-pulsar signals. However, as the number of labelled data increases, information about the persistence of the signal in ``Freq-Phase'' and ``Time-Phase'' regime become equally important. The individual score from each feature were combined using a Logistic regression model with ``L2'' regularization. This is marked as the combined model. Ideally, the combined model should be the best performing model. This holds true for our experiments with labelled candidates greater than five hundred. However, our combined model performs worse in the low labelled data regime (labelled candidates = 100) because the models trained on ``Pulse-Profile'', ``Time-Phase'' and ``Freq-Phase'' brings down the net average performance of the network. 
\begin{comment}

We refer the readers to the appendix for a full list of the model architecture and hyper-parameters used to train our models. 
\end{comment}

\subsection{Semi-supervised GAN}
\subsubsection{Model Architecture and Implementation Details}
There are three major components to an SGAN Network. A supervised discriminator, an unsupervised discriminator and an unsupervised generator.  The simplest implementation is to have a single discriminator with multiple output layers. The first output layer solves the unsupervised task and outputs if the data is REAL/FAKE and the second layer solves the supervised task and outputs if the signal is from a pulsar or not. The drawback of this approach is that when we pass unlabelled candidates from the generator, there is no supervised label associated with them. Hence, this creates  the need for an extra `FAKE' class label for the supervised classifier. In this paper, we follow the technique described in \citet{NIPS2016_6125} which removes the need for an extra class label. In this case, we built two separate models for the supervised and unsupervised task. Both models share the same feature extraction layers. However, the supervised model is  attached to a softmax activation function whereas the unsupervised model takes the output from the supervised model prior to the activation function and calculates a normalized sum of exponential outputs. This custom activation function for the unsupervised discriminator \emph{D(x)} forces the model to give a strong prediction for real samples and lower values for the generated fake samples
\begin{equation}
\label{unsupervised_equation}
    D(x) = \frac{Z(x)}{Z(x) + 1}, 
    \text{where} \, Z(x) = \Sigma_{k=1}^K exp[l_k (x)].
\end{equation}
\begin{figure*}

	\includegraphics[width=0.99\linewidth]{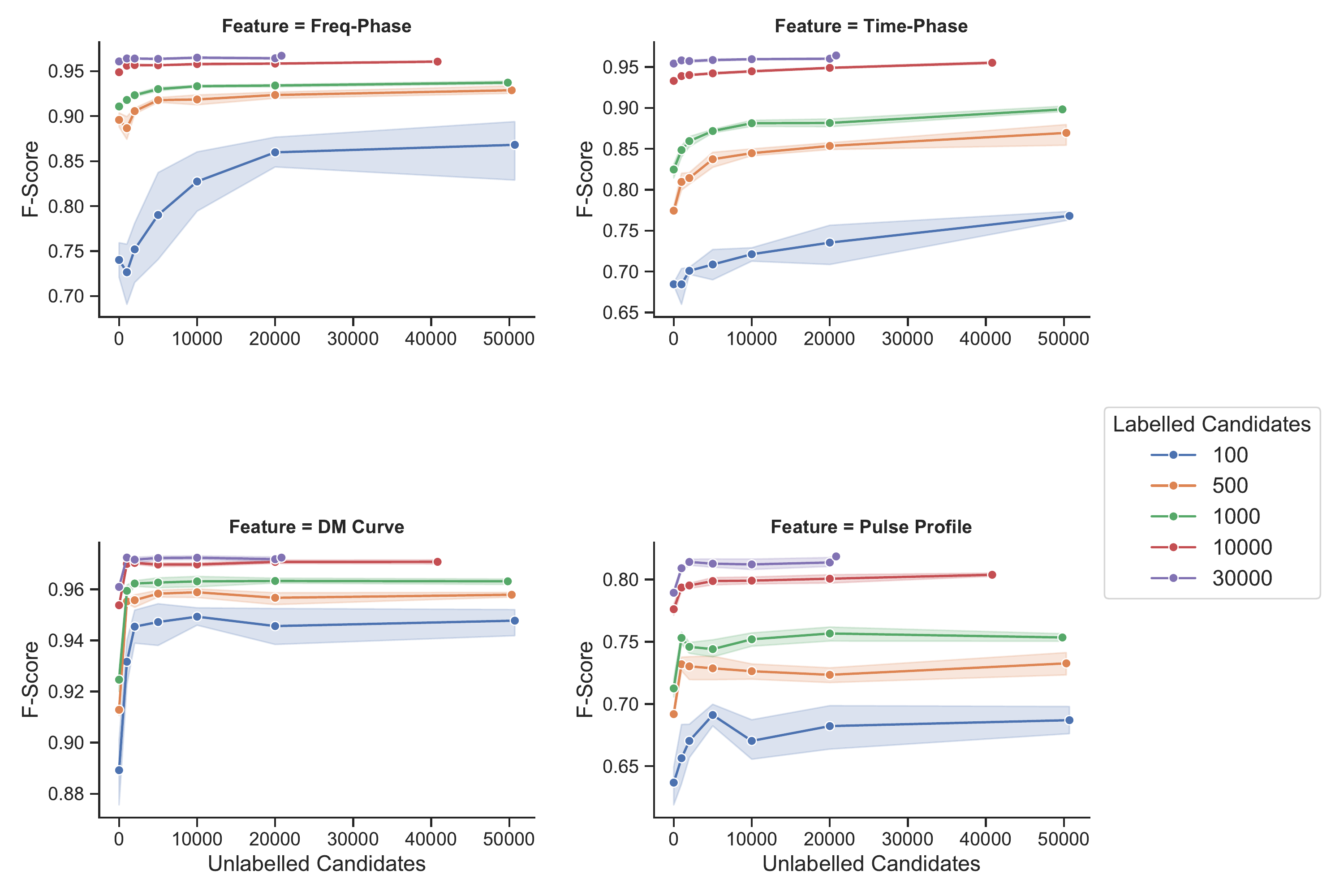}
	
	\caption{Mean F-score performance of each of the four features trained for 400 epochs with varying amounts of labelled and unlabelled samples. The interval around each line indicates the 95 \% confidence interval associated to each value. This figure demonstrates that unlabelled candidates can be leveraged to improve the overall performance of a machine learning algorithm. This effect is particularly dominant with labelled candidates $\leq$ 1000. The boost in performance of the ``Pulse-Profile'' and ``DM-Curve'' feature for labelled candidates $\geq$ 10000, was critical to improve the overall performance of the combined model. Similar improvements were also seen for all metrics  defined in section \ref{sec:metrics used}. These results can be found in table \ref{full_comparision_sgan_vs_supervised}.
	}
    \label{sgan_unlabelled_perf}
\end{figure*}

Our work is built on top of an open-source implementation of SGAN networks for MNIST digits\footnote{\url{https://machinelearningmastery.com/semi-supervised-generative-adversarial-network/}
}. We extensively modified the discriminator and generator architecture in order to get better results for our data. The discriminator architecture is similar to the CNN model used for the supervised baseline model. We obtained better results with larger convolutional kernels of size 7x7 compared to the 3x3 kernels that worked well for the supervised baseline models. The discriminator trained on the ``DM Curve'' and ``Pulse-Profile'' was a 1-D convolutional neural network with a convolutional kernel size of 7. Additionally, we used max-pooling to down-sample our images instead of stride convolutions. For the generator, we perform the transpose of convolutions in order to create images that are fed into the discriminator. Additionally, we also do batch normalization in order to speed up the training of the generator. The generator for the 1-D data was a multi-layer fully connected dense neural network. We use the tanh function as the activation function for the output layer of the generator. GANs can easily suffer from overconfidence. Therefore, as a regularization technique, we used soft and noisy labels while training. This means that if a candidate is real, instead of giving the label a value equals 1, we give a value in a range between 0.7-1.2 for the 2-D features and a value between 0.9-1 for the 1-D features. Around 5 per cent of the time, we intentionally flip the labels, we found that this helps to improve the overall performance. Our best performing model uses the Adam Optimizer \citep{2014arXiv1412.6980K} with a learning rate of $\eta = 0.0002$ and $\beta_1 = 0.5$.  

\begin{figure*}
\begin{center}
	\includegraphics[width=0.99\textwidth]{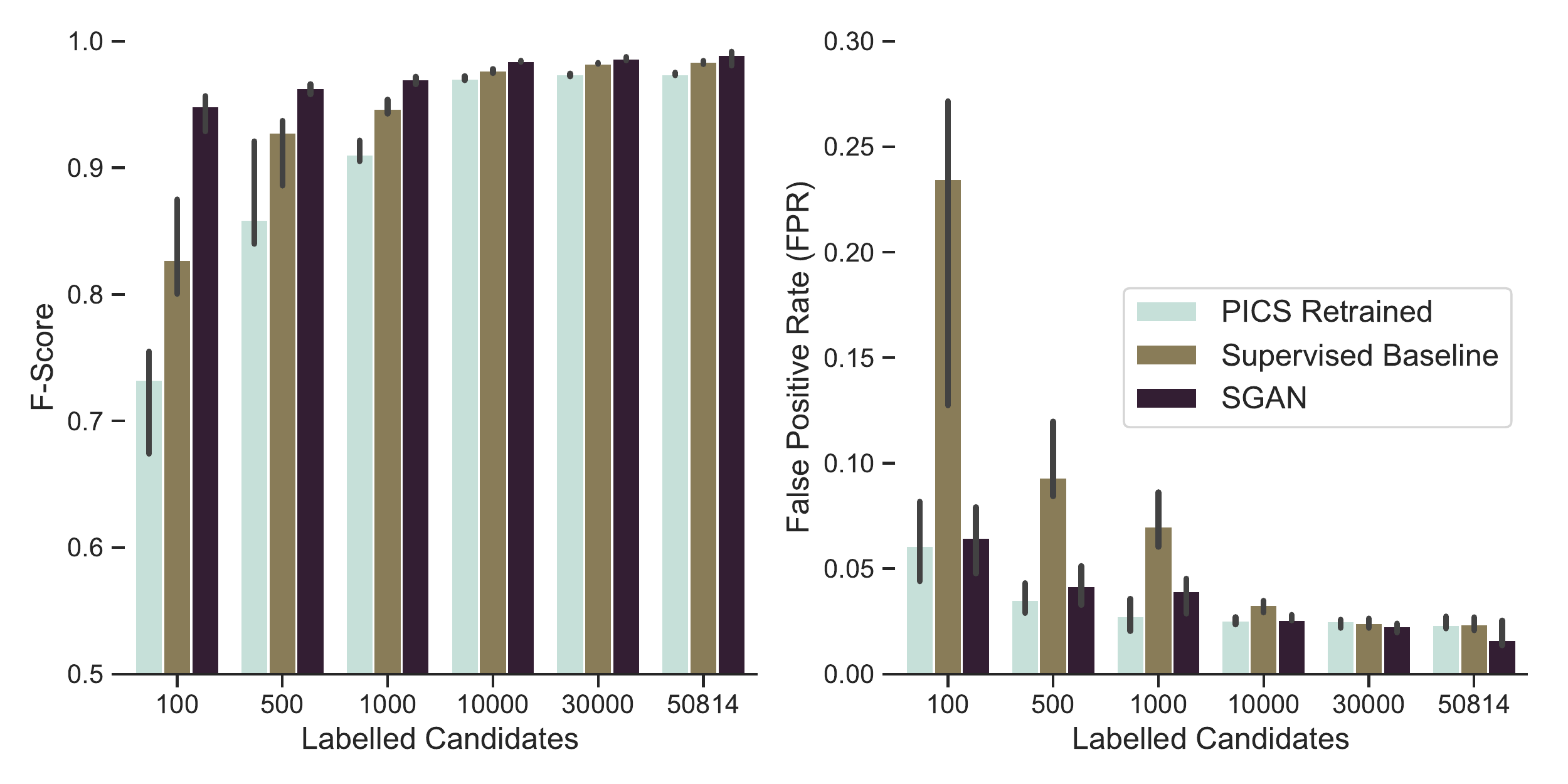}
\end{center}
\caption{Mean F-score performance (left) and False-Positive Rate (right) of the Combined Semi-Supervised GAN (SGAN), our  Supervised Baseline Model and the retrained version of Pulsar Image Classification System (PICS). The black vertical line indicates the 95\% confidence interval of our values. These results indicate that the SGAN architecture provides better classification results compared to both the supervised machine learning algorithms for all combinations of labelled data. For labelled candidates below 1000, the re-trained version of PICS has a lower false-positive rate compared to SGAN, however this comes at the cost of a significantly lower F-Score, thus making it a less desirable model. The difference in performance between the supervised baseline model and the retrained version of PICS can be partially attributed to the fact that PICS was not re-trained on thee same validation dataset. For labelled candidates below 1000, each of the supervised model ended up optimising for different metrics. Our supervised model resulted in a better recall rate and F-score at the cost of a worse false-positive rate and specificity score. We refer the readers to the appendix \ref{full_comparision_sgan_vs_supervised} for the scores across all metrics. The SGAN model provides the overall maximum gain when there are fewer labelled candidates (< 1000) available.}
\label{fig: sgan vs supervised}
\end{figure*}

\subsubsection{Effect of unlabelled data}
In order to test if the SGAN network can learn from unlabelled data, we split the training data-set into smaller groups ranging from 100, to 30,000 candidates similar to our supervised learning baseline model experiments. Similarly, the number of unlabelled candidates used while training was also varied from 0 to 20,000. We trained the SGAN network for 400 epochs in each configuration, saving only the model that produced the best results on the validation dataset. Since, the results are dependent on the subset of training samples used while training, we randomly selected five different combinations of labelled and unlabelled datasets and report the average values. The mean F-score values of SGAN trained on the all four features is  shown in Figure \ref{sgan_unlabelled_perf}. We observe that increasing the number of labelled candidates in the training set drastically improves the final performance of the network. In addition, we clearly see that unlabelled candidates also improve the overall performance of the network. This effect is particularly significant in the low labelled data regime. For example with 100 labelled candidates in the training set, the unlabelled data improved the F-score of the network by at least 6 \% for all features including an improvement of 12 \% on the network trained on the ``Freq-Phase'' feature. As the number of labelled candidates increase, we see that the semi-supervised classifier still provides better results. However, the performance boost provided by unlabelled candidates is significantly lower. We believe the reason for this is two-fold. With larger amount of labelled data, there is little room for improvement as the network has already learnt a good solution to solve the pulsar candidate identification problem. The second reason is that in order to fully utilise the strengths of the semi-supervised algorithm, we need to use a significantly large fraction of unlabelled candidates compared to the labelled candidates. Our final model which was trained on a much larger unlabelled candidate database has been described in Section \ref{sec:best performing}.

\subsection{Comparing SGAN with supervised models}
In this section, we compare the performance of our two-layer ensemble SGAN network to the ensemble standard supervised baseline algorithm described earlier as well as a re-trained version of the Pulsar Image Classification system (PICS) \citep{2014ApJ...781..117Z}. In all cases, the individual scores from each of the four features were combined using a Logistic regression model with L2 regularization. We trained all three networks with the same labelled candidates for each experiment. In addition, unlabelled candidates were also used to trained the SGAN model. The same validation dataset was used to tune hyperparameters for the supervised model and the SGAN model. We didn't use a validation set for re-training PICS. This was because there wasn't a provision to provide a validation dataset in the re-training script provided by \citep{2014ApJ...781..117Z}. We presume that PICS trained on minimising the overall training loss. We find that the ensemble SGAN outperforms the standard supervised baseline algorithm as well as re-trained version of PICS for all combinations of labelled datasets and based on all the metrics discussed in Section \ref{sec:metrics used} including higher accuracy, precision, recall rates and a lower false positive rate. For reasons of brevity we only show the mean F-score and False-Positive Rate (FPR) values in Figure \ref{fig: sgan vs supervised}. The full table comparing results of all the metrics can be found in Table \ref{full_comparision_sgan_vs_supervised} in the appendix.

\subsection{Best Performing Model}
\label{sec:best performing}
In this section, we describe our best performing model that was trained using the entire training set plus 265,172 unlabelled candidates. Results from five different training runs from the best performing semi-supervised and supervised models are summarised in Table \ref{tab:sgan_versus_supervised}. The confusion matrix of the predictions of this model on the test set is shown in Table \ref{tab:confusion matrix results}. Our best model achieved an overall F-score of 99.2 \%, recall rate of 99.7\% and a false positive rate of 1.63\%. Our best performing model has been merged into the HTRU-S Lowlat survey post-processing pipeline, and has already discovered eighteen new pulsars. These new pulsars had a detection significance ranging from 5.8 - 19 sigma. The SGAN network played a crucial role in discovering the lower detection significance pulsars as they are usually buried inside several non-pulsar candidates. A full list of these pulsars with their respective Spin-Period, DM and timing solutions will be the subject of a future publication.

\begin{table}
	\centering
	\caption{Results comparing the best performing combined supervised model trained on the entire training set against the best performing combined SGAN model trained on the entire training set plus 265,172 unlabelled candidates. Values reported in this table are the mean and standard deviation after repeating the training run five times.}
	\label{tab:sgan_versus_cnn}
	\begin{tabular}{lccc} % four columns, alignment for each
		\hline
		 & PICS Retrained & Supervised Baseline & SGAN\\
		\hline
		\vspace{0.1cm}
		Accuracy &$0.973\pm 0.001$ & $0.984\pm 0.001$ & $0.989\pm 0.004$ \\
		\vspace{0.1cm}
		Precision & $0.977 \pm 0.002$ & $0.977 \pm 0.002$ & $0.984\pm 0.004$ \\
		\vspace{0.1cm}
		Recall & $0.970\pm 0.001$ & $0.991\pm 0.001$ & $0.994\pm 0.004$ \\
		
		\vspace{0.1cm}
		
		F-score &$0.973\pm 0.001$ & $0.984\pm 0.001$ & $0.989\pm 0.004$ \\
		\vspace{0.1cm}
		Specificity & $0.976\pm 0.002$ & $0.976\pm 0.002$ & $0.983\pm 0.004$ \\
		\vspace{0.1cm}
		G-Mean & $0.973 \pm 0.001$ & $0.983\pm 0.001$ & $0.989\pm 0.004$ \\
		\vspace{0.1cm}
		FPR &$0.023\pm 0.002$ & $0.023\pm 0.002$ & $0.016 \pm 0.004$ \\
		%3 & 5 & 7 & 9\\
		\hline
		\label{tab:sgan_versus_supervised}
	\end{tabular}
\end{table}

\makegapedcells
\begin{table}
	\centering
	\caption{Normalised confusion matrix of the predictions of the best ensemble SGAN model on the test set.}
\begin{tabular}{cc|cc}
\multicolumn{2}{c}{}
            &   \multicolumn{2}{c}{Predicted} \\
    &       &   Pulsar &   Non-Pulsar              \\ 
    \cline{2-4}
\multirow{2}{*}{\rotatebox[origin=c]{90}{True}}
    & Pulsar   & 0.997   & 0.003                 \\
    & Non-Pulsar    & 0.014    & 0.986                \\ 
    \cline{2-4}
   
    \end{tabular}
    \label{tab:confusion matrix results}
 \end{table} 

\begin{figure*}
\begin{center}
	\includegraphics[width=0.85\textwidth]{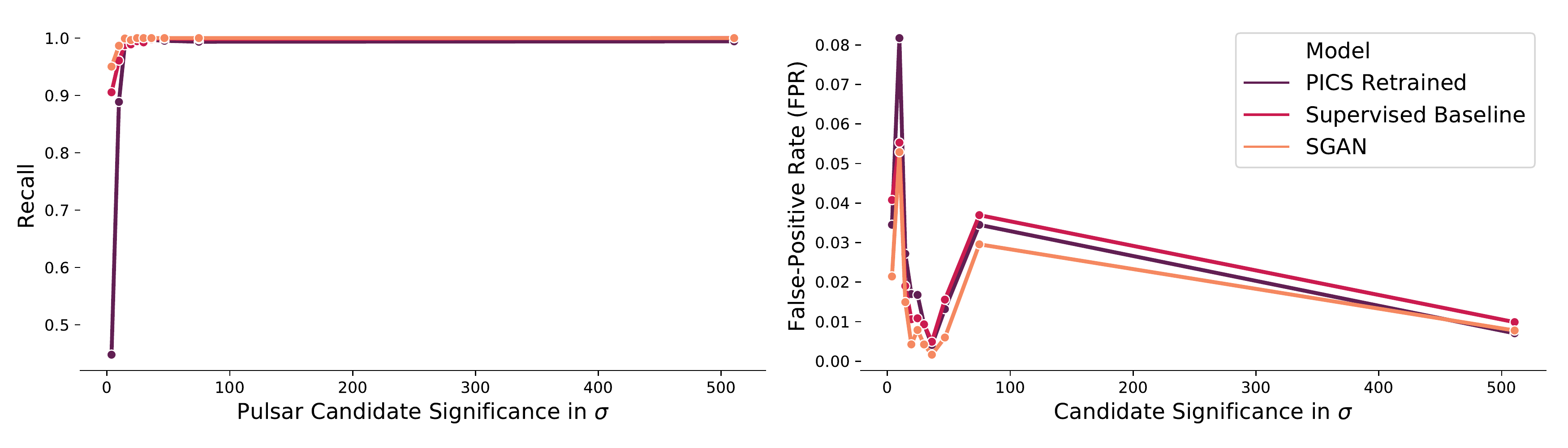}
\end{center}
\caption{Recall rate (left) and False-Positive rate (FPR) (right) from the best-performing models of the three networks across different candidate detection significance levels (SNR). For the x-axis, we divided the candidates in our test dataset into ten quantile regions based on their detection significance level with a similar number of candidates across each bin. As expected, all three models perform better when the significance level of the candidate is higher. The re-trained version of PICS suffers from a large performance loss (recall: 0.45) at lower candidate significance levels (0-7.4 sigma). The SGAN network also suffers a small performance loss, however it still does better compared to the other models. The FPR for all three networks are much more interesting. As expected, we see a large FPR when the detection significance is lower, followed by a lower FPR rate. However, we again see an increase in the FPR rate at high significance levels. This is mostly caused by bright broadband RFI signals which look like pulsars and are detected with very high significance levels. The FPR rate is highest for significance levels between 7.4-12.1 sigma. These are mostly weak pulsar-like signals that are caused white-noise lining up to look like pulsars. See table \ref{tab:sigma_vs_recall} for the performance of the network across other metrics.}
\label{fig:recall_fpr_snr}
\end{figure*}

\begin{figure*}
\begin{center}
	\includegraphics[width=0.85\textwidth]{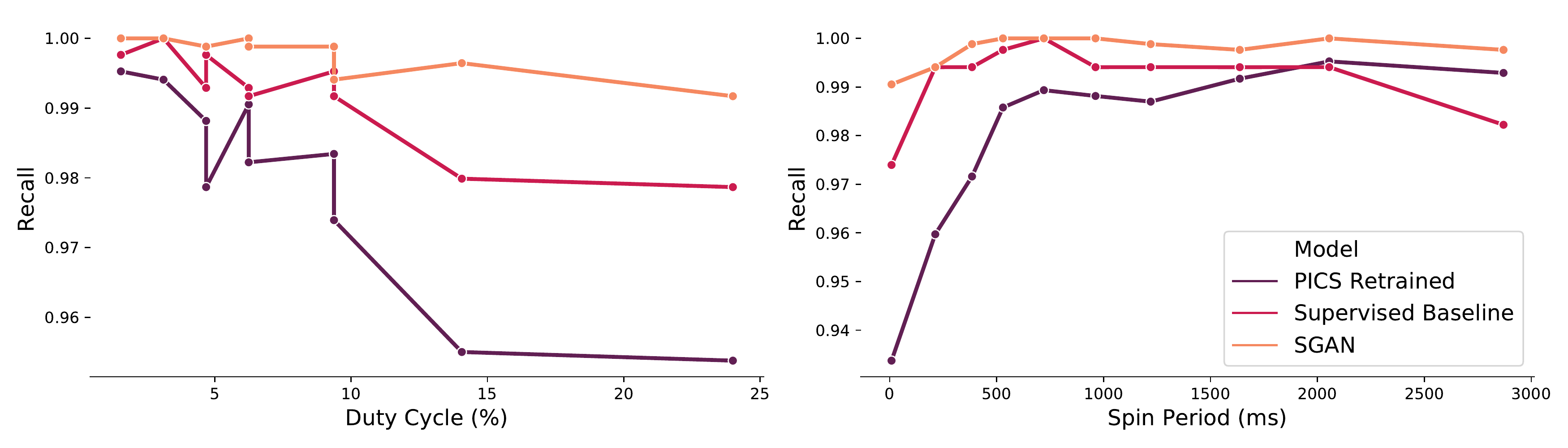}
\end{center}
\caption{Recall rate of all three networks across different pulsar duty cycle ranges (left) and Spin-Period (right). The x-axis of these plots were made by dividing the duty cycle and spin-period of all pulsars in our test dataset across ten quantile regions such that they have a similar number of pulsars in each bin. It is easier to spot narrow duty cycle pulsars by eye, we see a similar effect with our trained neural networks. The magnitude of difference however varies across each model and is significantly lower for the SGAN model. Our models also tend to do slightly better for slow pulsars. However, both these effects could be correlated as pulsars with a slower-spin period also tend to have a narrow duty cycle.}
\label{fig:duty cycle and spin p vs recall}
\end{figure*}

\subsection{Performance across Detection Significance, Duty cycle and Spin Period}
In this section, we briefly analyse the performance of the best model from all three networks with various pulsar parameters like detection significance, duty cycle and spin-period. We start by splitting all the candidates in our test dataset based on their detection significance into ten quantile regions such that each bin has a similar number of candidates. We then calculate the performance of the best performing model from each of the three networks described earlier in this bin range. This is shown in Figure \ref{fig:recall_fpr_snr}. As expected, we see an improvement in the recall rate of the network for higher detection significance. The performance drop for the re-trained version of PICS is drastic as it drops to a recall rate of 0.45 for pulsars with sigma values between 0 and 7.4. Both the supervised model and SGAN model do significantly better here scoring at 90.5 \% and 95.0\% respectively. The False Positive rate (FPR) is more interesting. As expected, we see a higher FPR at lower significance levels and visc versa. However, we see an increase in FPR rate for very high candidate significance levels (95.9 - 925.2 sigma). This was caused by bright broadband RFI signals that look very similar to pulsar signals. The FPR rate is highest for candidates with significance levels between 7.4 and 12.1 sigma. These are mostly caused by weak pulsar-like signals formed when white-noise lines up by chance. See table \ref{tab:sigma_vs_recall} for the performance of the network across additional metrics.

We also explored the performance of the neural network across the duty cycle of the pulsars in our test dataset. Similar to the previous experiment, we divide the candidates in the test set based on their duty cycle into ten quantile regions such that there are similar number of candidates in each bin. For a human, it is easier to spot pulsars with a narrow duty cycle. We see a similar trend in the performance of the neural networks as well. However, the performance drop is not as drastic as compared to the detection significance level. We also repeated this experiment across different spin-period ranges. Both of these are shown in Figure \ref{fig:duty cycle and spin p vs recall}. We see that pulsars with a slower spin-period appear to be slightly easier for the neural network to find. However, these two effects are highly correlated as slow spin-period pulsars also tend to have narrow duty cycle ranges.
\section{Discussion}
\subsection{Inference Speed}
 
  Since our software has been built using Keras and Tensorflow, the final trained models can be used to evaluate pulsar candidates both on a CPU or on any Nvidia GPU. The inference rate of our model, benchmarked on a single Nvidia Tesla P100 GPU, is $5.22 \pm 0.01$ ms using a batch size of 20,000. This makes our model particularly suitable for implementing in a blind survey like the HTRU-S Lowlat as the entire batch of 40 million candidates can be scored in $\approx$ 58 hours on a single GPU. Additionally, our architecture can be easily retrained and redeployed as more labelled data become available. Our software can be found on GitHub\footnote{\url{https://github.com/vishnubk/sgan}}. We also provide a Dockerfile which can be used to create a Docker image in order to ensure easy reproducibility of our results.
\subsection{Training Time and suitability for future pulsar surveys}

The training time for SGAN is considerably longer compared to training a standard supervised deep learning architecture like a CNN or an ANN. For example, our supervised baseline pipeline on average took less than an hour to finish training for 1000 epochs on a Nvidia Tesla P100 GPU whereas our final model from the SGAN architecture took about four hours to train for 400 epochs. The main reason for this is because we are now training two neural networks alternatively and using a considerably larger sample of data (unlabelled candidates) while training. Keras and Tensorflow currently support training on multiple GPUs which can help reduce the net training time. While the training time is still acceptable for our needs, this architecture may not be the best approach for online data processing where the model needs to be re-trained in quasi real time. We refer the readers to the work of \citet{2016MNRAS.459.1104L} that focuses more on speed rather than final classifier performance. The advantage of using our proposed architecture is higher performance because we can learn from unlabelled candidates. This is especially useful for RFI-rejection as such signals can have different signatures depending on the source (aircraft navigation, mobile phone, WiFi, satellites). Additionally, the RFI environment near a telescope is expected to change with the advancement of fifth generation wireless technology and therefore the ability to have a system that can adapt on relatively short timescales with high performance has huge value. Additionally, this technique also helps minimise the number of labelled candidates which saves human hours required to achieve satisfactory performance. The amount of unlabelled candidates that needs to be used while training depends on the number of labelled candidates available and is usually a trade-off between performance and training time. In our experiments, we achieved improved performance by having at least 5000 unlabelled candidates when the number of labelled candidates were below 1000. We observe that with 50,814 labels, we needed more than 200,000 unlabelled candidates to notice an improvement. It is also important to experiment with different amounts of unlabelled data as sometimes having more can make the model perform worse. We believe that using a static trained supervised model for classifying candidates from future pulsar surveys may not be the most optimal approach. Since labelling millions of candidates is not a scalable solution, we hope more attention goes into solving the pulsar candidate classification problem using a combination of labelled and unlabelled candidates.

\subsection{Future Work}
In this section, we briefly discuss some techniques that can be used to improve on our current models. 

\subsubsection{Improving the Supervised Baseline Model}
    The performance of our supervised baseline model in the regime of high labelled candidates can be improved using a much deeper convolutional neural network. Large Networks pretrained on ImageNet like VGG16 (\citet{2014arXiv1409.1556S}), InceptionV3 (\citet{2015arXiv151200567S}) and ResNet50 (\citet{2015arXiv151203385H}) among others can be used and their final few layers can be re-trained on a pulsar candidate dataset. This technique is called transfer learning and it has been successfully employed in various computer vision tasks including classifying Fast Radio Bursts (FRB) and RFI \citep{2019arXiv190206343A}. In order to have a fair comparison between such networks and SGAN, we propose using a similar deep architecture for the discriminator of SGAN and comparing their performance.

\subsubsection{Improving the SGAN Model}
We believe that the performance of our SGAN model can be improved further by using a technique called feature matching. For this, we change the loss function of the generator such that its goal changes from beating the discriminator to minimizing the statistical difference between real and generated images. We refer the readers to \citet{NIPS2016_6125} for a more detailed explanation of this technique. Another technique to improve the final semi-supervised classification accuracy is to use a Bad GAN. Instead of training towards a perfect generator, which produces images indistinguishable from real images, our goal in this architecture is to generate data that complements data produced from the discriminator. The drawback of this approach is that the quality of the generated images in general would be worse but this architecture has been shown to provide better classification results on the MNIST, SVHN and CIFAR-10 datasets \citep{2017arXiv170509783D}.

\section{Conclusion}
In this paper we use an ensemble Semi-Supervised Generative Adversarial (SGAN) framework to classify pulsar candidates in the HTRU-S Lowlat Survey. We demonstrate that this algorithm achieves an overall F-score of 99.2\% on our dataset and outperforms the standard supervised baseline algorithm and the re-trained version of PICS. The performance difference between both the techniques is significant in the low labelled-candidate regime. SGAN achieved a recall rate of 96.0\% with 100 labelled-candidates compared to 85.6\% from our supervised baseline model and 60.3 \% from the retrained version of PICS. The main advantage of our proposed network is the ability to leverage readily available unlabelled candidates for achieving better results. We believe this technique will be even more useful for future pulsar surveys as the number of pulsar candidates scale up and maintaining a large labelled dataset becomes increasingly challenging. Our architectures are frequency and telescope agnostic, therefore they can be in principle applied to other ongoing pulsar surveys. We additionally share our code, and a dockerfile to enable reproduciblity of our work.  

\section*{DATA AVAILABILITY}
The data underlying this article will be shared on reasonable request to the corresponding author.

\section*{Acknowledgements}
Observational data used in this work were made available by High Time Resolution Universe (HTRU) scientific collaboration. The Parkes Observatory, used in the collection of this data is part of the Australia Telescope National Facility which is funded by the Australian Government for operation as a National Facility managed by CSIRO. The data analysis were performed on the OzSTAR national supercomputing facilities at Swinburne University of Technology and the HERCULES computing cluster operated by the
Max Planck Computing \& Data Facility (MPCDF). OzSTAR is funded under Astronomy National Collaborative Research Infrastructure Strategy (NCRIS) Program via Astronomy Australia Ltd (AAL). We would like to thank members of the open-source community for maintaining packages that were directly used for our work including NumPy \citet{oliphant2006guide}, Matplotlib \citet{Hunter:2007}, Seaborn \citet{michael_waskom_2017_883859}, Scikit-learn \citet{scikit-learn}, Keras \citep{chollet2015keras} and Tensorflow \citep{tensorflow2015-whitepaper}.

%%%%%%%%%%%%%%%%%%%%%%%%%%%%%%%%%%%%%%%%%%%%%%%%%%



\bibliographystyle{mnras}
\bibliography{latest_draft} % if your bibtex file is called example.bib

%\begin{thebibliography}{99}
%\end{thebibliography}

%%%%%%%%%%%%%%%%%%%%%%%%%%%%%%%%%%%%%%%%%%%%%%%%%%

%%%%%%%%%%%%%%%%% APPENDICES %%%%%%%%%%%%%%%%%%%%%

\appendix

\section{Detailed Performance Comparison of SGAN, Re-trained PICS and Supervised Baseline Model.}

\begin{table*}
\caption{Detailed performance comparison of Semi-Supervised GAN and the our Ensemble Supervised Machine Learning Baseline Model for varying amounts of labelled data. Each row represents the mean performance of the network after averaging across five different batches of labelled data.}
\renewcommand\arraystretch{1.25} % increase spacing between rows
\resizebox{\textwidth}{!}{%
\begin{tabular}{lllrrrrrrrrrrrrrr}
\toprule
      &        &      & \multicolumn{2}{l}{Accuracy} & \multicolumn{2}{l}{F-Score} & \multicolumn{2}{l}{FPR} & \multicolumn{2}{l}{G-Mean} & \multicolumn{2}{l}{Precision} & \multicolumn{2}{l}{Recall} & \multicolumn{2}{l}{Specificity} \\
      &        &      &    median &       std &    median &       std &    median &       std &    median &       std &    median &       std &    median &       std &      median &       std \\
Labelled Samples & Unlabelled Samples & neural\_net &           &           &           &           &           &           &           &           &           &           &           &           &             &           \\
\midrule
100   & 0      & PICS Retrained &  0.778114 &  0.021186 &  0.732734 &  0.035151 &  0.060684 &  0.014769 &  0.759390 &  0.028041 &  0.905080 &  0.015769 &  0.603222 &  0.050334 &    0.939316 &  0.014769 \\
      &        & Supervised Baseline &  0.811978 &  0.030802 &  0.827461 &  0.027541 &  0.234638 &  0.057758 &  0.807474 &  0.031681 &  0.784844 &  0.043643 &  0.856594 &  0.030882 &    0.765362 &  0.057758 \\
      & 5000   & SGAN &  0.947764 &  0.011199 &  0.948811 &  0.011012 &  0.064780 &  0.012405 &  0.947577 &  0.011209 &  0.937786 &  0.011890 &  0.959910 &  0.011033 &    0.935220 &  0.012405 \\
500   & 0      & PICS Retrained &  0.870921 &  0.026020 &  0.858840 &  0.030660 &  0.035343 &  0.005588 &  0.866737 &  0.027780 &  0.957283 &  0.007754 &  0.783908 &  0.048144 &    0.964657 &  0.005588 \\
      &        & Supervised Baseline &  0.925943 &  0.019918 &  0.927882 &  0.020159 &  0.093265 &  0.014035 &  0.925586 &  0.019763 &  0.911531 &  0.014496 &  0.944830 &  0.027332 &    0.906735 &  0.014035 \\
      & 10000  & SGAN &  0.962169 &  0.003598 &  0.962963 &  0.003699 &  0.041726 &  0.006985 &  0.961964 &  0.003566 &  0.959504 &  0.006289 &  0.972087 &  0.009891 &    0.958274 &  0.006985 \\
1000  & 0      & PICS Retrained &  0.915647 &  0.005872 &  0.910673 &  0.006351 &  0.027532 &  0.006123 &  0.913985 &  0.005989 &  0.969684 &  0.006642 &  0.859123 &  0.009233 &    0.972468 &  0.006123 \\
      &        & Supervised Baseline &  0.945686 &  0.004898 &  0.947000 &  0.004645 &  0.070020 &  0.009455 &  0.945394 &  0.004990 &  0.932637 &  0.008242 &  0.963188 &  0.005629 &    0.929980 &  0.009455 \\
      & 20000  & SGAN &  0.969537 &  0.002386 &  0.970052 &  0.002626 &  0.039535 &  0.007066 &  0.969419 &  0.002255 &  0.961790 &  0.006225 &  0.978456 &  0.011495 &    0.960465 &  0.007066 \\
10000 & 0      & PICS Retrained &  0.970576 &  0.001187 &  0.970700 &  0.001225 &  0.025436 &  0.001286 &  0.970601 &  0.001182 &  0.974780 &  0.001203 &  0.966842 &  0.002887 &    0.974564 &  0.001286 \\
      &        & Supervised Baseline &  0.976527 &  0.001230 &  0.976850 &  0.001208 &  0.032867 &  0.002169 &  0.976462 &  0.001232 &  0.968276 &  0.002017 &  0.986043 &  0.002044 &    0.967133 &  0.002169 \\
      & 20000  & SGAN &  0.983989 &  0.000400 &  0.984271 &  0.000387 &  0.025817 &  0.001091 &  0.983861 &  0.000407 &  0.975074 &  0.001011 &  0.994099 &  0.000755 &    0.974183 &  0.001091 \\
30000 & 0      & PICS Retrained &  0.973646 &  0.000919 &  0.973827 &  0.000905 &  0.025055 &  0.001558 &  0.973656 &  0.000921 &  0.975291 &  0.001504 &  0.971525 &  0.001054 &    0.974945 &  0.001558 \\
      &        & Supervised Baseline &  0.982430 &  0.000326 &  0.982686 &  0.000305 &  0.024102 &  0.001742 &  0.982354 &  0.000340 &  0.976596 &  0.001623 &  0.988854 &  0.001309 &    0.975898 &  0.001742 \\
      & 20000  & SGAN &  0.985973 &  0.001134 &  0.986206 &  0.001106 &  0.022673 &  0.001933 &  0.985863 &  0.001146 &  0.978075 &  0.001838 &  0.994005 &  0.000566 &    0.977327 &  0.001933 \\
50814 & 0      & PICS Retrained &  0.973693 &  0.000928 &  0.973829 &  0.000904 &  0.023245 &  0.002171 &  0.973714 &  0.000933 &  0.976996 &  0.002072 &  0.970963 &  0.001352 &    0.976755 &  0.002171 \\
      &        & Supervised Baseline &  0.983847 &  0.001124 &  0.984099 &  0.001093 &  0.023721 &  0.002387 &  0.983755 &  0.001138 &  0.977013 &  0.002252 &  0.991289 &  0.001282 &    0.976279 &  0.002387 \\
      & 265172 & SGAN &  0.989468 &  0.004313 &  0.989614 &  0.004251 &  0.016290 &  0.004592 &  0.989403 &  0.004317 &  0.984159 &  0.004456 &  0.994661 &  0.004115 &    0.983710 &  0.004592 \\
\bottomrule
\end{tabular}

}
\label{full_comparision_sgan_vs_supervised}
\end{table*}

\begin{table*}

\caption{Performance comparison of SGAN, retrained version of PICS and our Supervised Baseline Model for different ranges of detection sigma values.}
\renewcommand\arraystretch{1.15} % increase spacing between rows
\resizebox{\textwidth}{!}{%
\begin{tabular}{llrrrrrrrr}
\toprule
    &                     &       FPR &  Total Candidates &  Precision &  Sigma Avg &  Sigma Max &  Sigma Min &    Recall &  Specificity \\
Percentile & Model &           &                   &            &               &               &               &           &              \\
\midrule
10  & PICS Retrained &  0.034501 &              2114 &   0.576923 &         3.700 &          7.40 &          0.00 &  0.447761 &     0.965499 \\
    & SGAN &  0.021432 &              2114 &   0.823276 &         3.700 &          7.40 &          0.00 &  0.950249 &     0.978568 \\
    & Supervised Baseline &  0.040774 &              2114 &   0.700000 &         3.700 &          7.40 &          0.00 &  0.905473 &     0.959226 \\
20  & PICS Retrained &  0.081731 &              2107 &   0.943381 &         9.750 &         12.10 &          7.40 &  0.888627 &     0.918269 \\
    & SGAN &  0.052885 &              2107 &   0.966206 &         9.750 &         12.10 &          7.40 &  0.986667 &     0.947115 \\
    & Supervised Baseline &  0.055288 &              2107 &   0.963808 &         9.750 &         12.10 &          7.40 &  0.960784 &     0.944712 \\
30  & PICS Retrained &  0.027174 &              2107 &   0.985465 &        14.550 &         17.00 &         12.10 &  0.989059 &     0.972826 \\
    & SGAN &  0.014946 &              2107 &   0.992035 &        14.550 &         17.00 &         12.10 &  0.999271 &     0.985054 \\
    & Supervised Baseline &  0.019022 &              2107 &   0.989766 &        14.550 &         17.00 &         12.10 &  0.987600 &     0.980978 \\
40  & PICS Retrained &  0.016985 &              2117 &   0.986498 &        19.450 &         21.90 &         17.00 &  0.994894 &     0.983015 \\
    & SGAN &  0.004246 &              2117 &   0.996596 &        19.450 &         21.90 &         17.00 &  0.996596 &     0.995754 \\
    & Supervised Baseline &  0.010616 &              2117 &   0.991468 &        19.450 &         21.90 &         17.00 &  0.988936 &     0.989384 \\
50  & PICS Retrained &  0.016765 &              2110 &   0.984657 &        24.500 &         27.10 &         21.90 &  0.995438 &     0.983235 \\
    & SGAN &  0.007890 &              2110 &   0.992754 &        24.500 &         27.10 &         21.90 &  1.000000 &     0.992110 \\
    & Supervised Baseline &  0.010848 &              2110 &   0.990009 &        24.500 &         27.10 &         21.90 &  0.994526 &     0.989152 \\
60  & PICS Retrained &  0.009314 &              2116 &   0.988285 &        29.800 &         32.50 &         27.10 &  0.992513 &     0.990686 \\
    & SGAN &  0.004234 &              2116 &   0.994681 &        29.800 &         32.50 &         27.10 &  1.000000 &     0.995766 \\
    & Supervised Baseline &  0.009314 &              2116 &   0.988285 &        29.800 &         32.50 &         27.10 &  0.992513 &     0.990686 \\
70  & PICS Retrained &  0.004095 &              2149 &   0.994624 &        36.185 &         39.87 &         32.50 &  0.996767 &     0.995905 \\
    & SGAN &  0.001638 &              2149 &   0.997849 &        36.185 &         39.87 &         32.50 &  1.000000 &     0.998362 \\
    & Supervised Baseline &  0.004914 &              2149 &   0.993562 &        36.185 &         39.87 &         32.50 &  0.997845 &     0.995086 \\
80  & PICS Retrained &  0.013174 &              2111 &   0.991413 &        46.785 &         53.70 &         39.87 &  0.995298 &     0.986826 \\
    & SGAN &  0.005988 &              2111 &   0.996097 &        46.785 &         53.70 &         39.87 &  1.000000 &     0.994012 \\
    & Supervised Baseline &  0.015569 &              2111 &   0.989907 &        46.785 &         53.70 &         39.87 &  0.999216 &     0.984431 \\
90  & PICS Retrained &  0.034483 &              2123 &   0.991860 &        74.795 &         95.89 &         53.70 &  0.993593 &     0.965517 \\
    & SGAN &  0.029557 &              2123 &   0.993060 &        74.795 &         95.89 &         53.70 &  1.000000 &     0.970443 \\
    & Supervised Baseline &  0.036946 &              2123 &   0.991324 &        74.795 &         95.89 &         53.70 &  0.998253 &     0.963054 \\
100 & PICS Retrained &  0.007057 &              2118 &   0.985856 &       510.545 &        925.20 &         95.89 &  0.994294 &     0.992943 \\
    & SGAN &  0.007763 &              2118 &   0.984551 &       510.545 &        925.20 &         95.89 &  1.000000 &     0.992237 \\
    & Supervised Baseline &  0.009880 &              2118 &   0.980420 &       510.545 &        925.20 &         95.89 &  1.000000 &     0.990120 \\
\bottomrule
\end{tabular}
}
\label{tab:sigma_vs_recall}
%\label{full_comparision_sgan_vs_supervised}
\end{table*}

%%%%%%%%%%%%%%%%%%%%%%%%%%%%%%%%%%%%%%%%%%%%%%%%%%

% Don't change these lines
\bsp	% typesetting comment
\label{lastpage}
\end{document}